\documentclass[prc,twocolumn,nofootinbib]{revtex4}

\usepackage{graphicx}
\usepackage[usenames,dvipsnames]{color}
\usepackage{amsmath,amssymb,bbold}
\usepackage{hyperref}
\usepackage{comment}
\usepackage{graphicx}
\usepackage{dcolumn}
\usepackage{bm}
\usepackage{amsmath}
\usepackage{bm}
\usepackage{ulem}
\usepackage{soul}

\begin{document}

\title{Thermoelectric properties of the (an-)isotropic QGP in magnetic fields}

\author{He-Xia Zhang}
\email{zhanghexia@mails.ccnu.edu.cn}
\affiliation{Key Laboratory of Quark \& Lepton Physics (MOE) and Institute of 
	Particle Physics, Central China Normal University, Wuhan 430079, China}

\author{Jin-Wen Kang}
\affiliation{Key Laboratory of Quark \& Lepton Physics (MOE) and Institute of 
	Particle Physics, Central China Normal University, Wuhan 430079, China}

 \author{Ben-Wei Zhang}
\email{bwzhang@mail.ccnu.edu.cn}
\affiliation{Key Laboratory of Quark \& Lepton Physics (MOE) and Institute of  Particle Physics, Central China Normal University, Wuhan 430079, China}
\affiliation{Institute of Quantum Matter, South China Normal University, Guangzhou 510006, China}

\begin{abstract}
  The Seebeck effect and the  Nernst effect, which reflect the appearance of  electric fields along $x$-axis and along $y$-axis ($E_{x}$ and $E_{y}$), respectively, induced by  the thermal  gradient along $x$-axis,
	are studied in the  QGP at an external   magnetic field along  $z$-axis. We calculate the associated  Seebeck coefficient ($S_{xx}$) and Nernst signal ($N$) using  the relativistic Boltzmann   equation under the  relaxation time approximation. 	In an  isotropic QGP, the influences of magnetic field ($B$) and quark chemical potential ($\mu_{q}$) on these thermoelectric transport coefficients are investigated.
	In the presence (absence) of weak magnetic field, we find $S_{xx}$ for a fixed  $\mu_{q}$ is negative (positive) in sign, indicating that the dominant carriers for converting heat gradient to electric field are negatively (positively) charged quarks. The absolute value of  $S_{xx}$  decreases with increasing temperature.
 	Unlike $S_{xx}$,  the sign of $N$ is independent of charge carrier type,   and  its thermal behavior displays a peak structure.
 	In the presence of strong magnetic field, due to the Landau quantization of  
 	  transverse motion of   (anti-)quarks perpendicular to magnetic   field, only the longitudinal Seebeck coefficient ($S_{zz}$) exists. Our results show that  the value of  $S_{zz}$ at a fixed $\mu_{q}$ in the  lowest Landau level (LLL)  approximation always remains positive. 
   Within the effect of high Landau levels,  $S_{zz}$  exhibits  a thermal  structure  similar to that in the  LLL approximation.
   As   the Landau level increases further,  $S_{zz}$ decreases and even its sign  changes from positive to negative.   The computations of these thermoelectric transport coefficients are also extended to  a medium with  momentum-anisotropy induced by initial spatial expansion as well as strong magnetic field.
\end{abstract}
\maketitle

\section{Introduction}
\label{sec:intro}

Quantum chromodynamics (QCD) is the fundamental theory of the strong interaction, and  the Lattice QCD calculations have predicted a crossover phase transition from the  hadronic matter to the   quark-gluon plasma (QGP)  can be realized with the increase of temperature at the small or vanishing baryon chemical potential~\cite{Cheng:2006qk,Aoki:2006br}. Heavy-ion collisions (HICs) experiments with very large center-of-mass colliding energies, e.g., the relativistic heavy-ion collision (RHIC) at BNL and the large hadron collision (LHC) at CERN also provide an opportunity to deeply  exploring  the phase structure and the transport properties of strongly interacting matter at the extreme conditions. 
In the  non-central HICs,  the presence of an enormous  magnetic  field in the direction perpendicular to the reaction plane is  expected~\cite{Rafelski:1975rf}. The theoretical estimate of field strength at primary stage 
of collisions can reach $ eB\sim m_{\pi}^{2}\sim 0.02 $ $\mathrm{GeV^{2}} $
for RHIC and $eB\sim15m_{\pi}^{2}\sim 0.3$ $\mathrm{GeV^{2}}$ for LHC~\cite{magnetic, Bzdak:2011yy,Deng:2012pc,Skokov:2009qp,Kharzeev:2007jp,delay}. And  this   magnetic field  can  persist long-lived due to the presence of electrical conductivity of medium~\cite{delay,Tuchin:2010gx,delay2}.
In the past  years, a variety of novel insights of strongly interacting matter induced by strong magnetic background field have sparkled  considerable research, such as chiral magnetic effect~\cite{Fukushima,chiral-magentic1,chiral-magnetic2}, the chiral magnetic  wave~\cite{chiral-wave1,chiral-wave3}, inverse 
magnetic catalysis~\cite{inverse1,inverse2,inverse3,inverse4,inverse5,inverse6,inverse7}, and the heavy quark transport~\cite{heavyquark1,heavyquark2,heavyquark3,heavyquark4,Kurian:2019nna,heavyquark6,heavyquark7}, ect.  Thus, investigating  the magnetic field-induced phenomenological consequences and the  effect of magnetic field on transport properties can provide a  comprehensive understanding of the complex QCD matter.

Transport coefficients, characterizing the dynamical evolution of system, play a crucial role to probe the  strongly interacting matter.  Recently, numerous works have been devoted to studying the effects of magnetic field on transport coefficients in QCD matter. Due to the  uncertainty  of the realistic magnitude of  magnetic field produced in the initial stage of HICs,  we only consider that the magnetic field is constant and homogeneous in present work. 
In the presence of  a  magnetic field  oriented along $z$-axis 
the Landau gauge $A_\mu=(0,0,-Bx,0)$ ($B$ is the strength of an artificial magnetic field) is chosen as usual, where $A_\mu$ denotes the electromagnetic  potential. 
By solving the Dirac equation of motion,  the dispersion relation for light  (anti-)quarks in the  QGP is obtained  quantum-mechanically as~\cite{Gusynin:1995nb,Akhiezer,Andersen:2014xxa} $\epsilon_{f,l}=\sqrt{p_{z}^2+m_{f}^2+2l|q_{f}eB|}$ 
($l=0,1,\dots$ are the quantum  numbers of the Landau energy levels;  $m_{f}$ is the current mass for $f$-th flavor  (anti-)quarks; $p_{z}$ is the momentum of charged particle along the  direction of magnetic field; $q_{f}e$ is the electric charge of $f$-th flavor quarks.). 
In the literature, 
the ranges of magnetic field can be roughly categorized into three scenarios: the weak magnetic case or classical case, the strong magnetic field case or higher  Landau levels (hLLs) case, the strong magnetic field limit case or lowest Landau level (LLL) case, which are implemented by different scale hierarchies.
In the weak  field case within the regime $g^2T^2\ll eB$ ($g$ is the QCD charge, $g\ll1$), the temperature acts as the dominant scale in the presence of magnetic field, the  quantum effect due to Landau quantization is not included,  the magnetic field effect  can be seen at the classical level in the  so-called cyclotron motion of charged particle.
Furthermore,  we reasonably assume that the scattering mechanism of partons and thermodynamics   in the  weakly magnetized medium are unaffected  by the presence of magnetic field.  
With the increase of $B$, in the strong magnetic field case within the regime $eB\gg g^2T^2$ introduced  in Ref.~\cite{electrical3}, the quantum effect  is increasingly obvious, the Landau quantization of  the  cyclotron motion  is seen  at the quantum level  and  the kinetic energy of  quarks gets discreted  into Landau levels.  As the magnetic field increases further, in the extreme magnetic field limit with the scale hierarchy $eB\gg T^2$,  the magnitude of magnetic field is sufficiently greater than other energy scales in the thermal medium, the transition from the LLL  to the hLLs  requires a large energy gap $~\sqrt{eB}$ to excite.  Consequently,  the contribution from  the hLLs  are neglected due to the  suppression of the Boltzmann factor  $\sim e^{-\sqrt{2leB}/T}$,  only  keeps the contributions from the LLL.
In the context of weak field,  electrical  conductivity of QGP recently have been computed using the  kinetic theory~\cite{bfeng} and quasi-particle models~\cite{Thakur:2019bnf,Das:2019ppb}. And the  systematic studies of shear viscosity at weak field  have been done within perturbative QCD in leading log~\cite{Li:2017tgi}.
These observables   have  also  been investigated in hadronic matter  at weak  magnetic field within
hadron resonance gas (HRG) model~\cite{Arpan,Das:2019pqd}.
At the strong magnetic field  within the LLL approximation,   electrical conductivity along the direction of magnetic field  in the  QGP   has been  estimated using diagrammatic method~\cite{electrical2},  perturbative QCD approach~\cite{Hattori:2016lqx}, and  effective quasi-particle model~\cite{Kurian:2017yxj}. 
In  the LLL approximation, the effect of magnetic field on other observables, such as
viscosities~\cite{transport3,transport2,eta3,bulk-viscosity1,Kurian:2018dbn}, heavy quark complex potential~\cite{Singh:2017nfa},
diffusion coefficients of heavy quark~\cite{heavyquark1,Kurian:2019nna}, heavy quark collisional energy loss~\cite{Singh:2020fsj},  the properties of quarkonium states~\cite{Hasan:2018kvx} and jet quenching parameter~\cite{jet quenching} also have been studied.
Furthermore, the effect of  hLLs on various transport coefficients has been  investigated   recently in Refs.~\cite{electrical3,electrical5,Kurian:2018dbn,Kurian:2019nna}.

Besides aforementioned common transport coefficients, some attention recently  has been turned to the studies of electromagnetic and thermoelectric  effects  such as the Hall effect, the Seebeck effect and  the Nernst effect, which are also fundamental  to understand the  electrical transport properties of QCD matter.
The  Hall effect describes the generation of a transverse electric field in an electric current-carrying conductor when a uniform magnetic field  perpendicular to the direction of current is applied, which is usually studied in solid materials.
In  the hot QCD matter, due to the  significant initial velocity of charged particles along the beam direction  is  perpendicular to the produced magnetic field created in non-central HICs, the Lorentz  force can result in an electric current normal to both the initial velocity of charged particles  and the magnetic field, which indicates the investigation of  the  Hall effect  in QCD matter is reasonable. Actually, the conductivity associated this  Hall current,  the Hall conductivity, in the baryon rich hadronic matter as well as the   QGP    has been estimated  already using the kinetic theory~\cite{Arpan,bfeng,Das:2019ppb}. The Hall component of shear viscosity has also recently  been studied in holographic model~\cite{Hoyos:2019pyz}.
Furthermore, in the presence of  magnetic field, a current of charge carriers can  be deflected,  whether  it is an electrical current or a thermal current. 
When a temperature gradient along $x$-axis ($\nabla_{x}T$) exists in a conducting medium, a corresponding electric field ($E_{x}$) can arise.
By applying an external magnetic field  along $z$-axis, the thermal current of charge carriers generated by  the  temperature gradient along $x$-axis can be deflected to $y$-axis, resulting in an  electric field along $y$-axis ($E_{y}$). The appearance of $E_x$ and $E_y$ due to the  thermal gradient along $x$-axis is called  the  Seebeck effect and the  Nernst effect, respectively. 
Accordingly, the proportionality constants, $\frac{E_{x}}{\nabla_{x}T}$ and $\frac{E_{y}}{\nabla_{x}T}$, in  zero-current condition  are called  the Seebeck coefficient ($S_{xx}$) and the  Nernst signal ($N$ or $S_{xy}$). 
The strength of   $S_{xx}$ and  $N$ reflects the  efficiency of the thermoelectric materials regarding the conversion of heat into electrical power. In condensed matter physics, the Seebeck effect  and the  Nernst effect have been studied in various solid state matters, such as  semiconductors~\cite{metals}, Bismuth~\cite{Bismuth}, graphene~\cite{graphene} and Weyl semimetal~\cite{Mandal:2020fmt,Lundgren:2014hra}.
The Seebeck  coefficient  and  the Nernst coefficient  have also been  estimated in  hot and dense hadronic matter at zero magnetic field~\cite{Bhatt:2018ncr} as well as at nonzero magnetic field~\cite{Das:2020beh}. The Nernst effect in a strongly correlated system at finite magnetic field has also been  studied by the gauge gravity duality~\cite{Kim:2015wba}.  Very recently, the Seebeck  coefficient of  QGP in the  LLL approximation has been computed  using quasi-particle model~\cite{Dey:2020sbm}.
To the best of our knowledge, there is no estimation of  the Seebeck effect and the  Nernst effect for  QGP in the  weak magnetic field case.  Hence, this provides one of  main motivations for the phenomenological research in this paper.
Conventionally,  an ideal assumption that the  constituents of  QGP or of   hadronic matter are isotropic in momentum space has been  employed in most existing estimations of thermoelectric coefficients. However, due to  the  geometry of primary fireball generated in HICs is  asymmetric, the different pressure gradients along different directions make  the  expansion along the  beam direction  (denoted by $\parallel$)  more rapid  than that along the directions perpendicular to beam direction (denoted by $\perp$), i.e. $p_{\parallel}\ll p_{\perp}$~\cite{Dumitru:2009ni,Srivastava:2016igg}.
The presence of   shear viscosity  also can contribute to such momentum-space anisotropy, thus the anisotropy  can survive a long time~\cite{Dumitru:2009ni}.
There are some studies to explore the influence of momentum-anisotropy  induced by the initial spatial expansion on various transport  coefficients~\cite{Srivastava:2016igg},
collective excitations of hot QCD medium~\cite{Kumar:2017bja} and quarkonium bound state~\cite{Margotta:2011ta,Dumitru:2009ni}. Apart from  the initial spatial expansion-driven momentum anisotropy, the momentum anisotropy also can be induced by  strong magnetic field. As mentioned earlier, in the  strong magnetic field  the motion of charged particles  perpendicular to the direction of magnetic field (denoted by $\perp$) can be quantum-mechanically  restricted due to Landau quantization,  the dynamic motion is mainly  along  the direction of magnetic field (denoted by $\parallel$),  i.e., $p_{\perp}\ll p_{\parallel}$. Recently, the influence of this  anisotropy  induced by strong field  on common  transport coefficients  has  also been analyzed  in Ref.~\cite{anisotropic-B}.  
Hence, it is also of interest to study   how the momentum anisotropy induced by initial spatial expansion and  strong magnetic field affects the quantitative and qualitative features of  thermoelectric  coefficients in the QGP.

In present work, we first calculate the  Seebeck coefficient ($S_{xx}$) and the Nernst signal ($N$) in  an (an-)isotropic QGP  at the weak  magnetic field case within the hierarchy of scale $eB\ll g^2T^2$.
Next, we calculate   the longitudinal Seebeck coefficient  ($S_{zz}$) in the (an-)isotropic QGP under the  LLL approximation within the  hierarchy of scales $\alpha_{s}eB\ll T^2\ll eB$ introduced by ~K. Fukushima $et ~al$~\cite{heavyquark1}.  In the strong magnetic field,  the Debye mass from quark-loops  is propotional to $\sqrt{\alpha_{s,B}eB}$ ($\alpha_{s,B}$ is magnetic field-dependent QCD running coupling constant)~\cite{Rath:2017fdv,Ferrer:2014qka,Ayala:2018wux}.
The   first inequality indicates that the self-energy corrections to hard LLL (anti-)quarks  and gluons can be reasonably neglected in leading order of  perturbative QCD calculation.
Different to the traditional binary scatterings,  due to the dimensional reduction of (anti-)quarks in the presence  of strong  magnetic field,  the   dimensional mixmatch between (anti-)quarks and gluons  leads to the novel scattering process, i.e., the   quark-antiquark pair to gluon is possible ~\cite{Hattori:2016lqx,Tuchin:2010gx}.
In the LLL approximation,  we  consider two  kind of  scattering processes, namely,  quark-antiquark pair to gluon  $2\rightarrow1$  process  and   usual quark-antiquark $t$-channel $2\rightarrow2$ process, where the small current (anti-)quark mass  cannot be ignored because the scatterings  are forbidden in  massless limit according to the chirality conservation~\cite{chirality}. 
When the magnetic field is not so large within the regime~$g^2T^2\ll eB$, the contribution from hLLs needs to be considered in the computation of $S_{zz}$.   Note that in this work  the magnetic field is  regarded as an external degree of freedom, namely, we neglect the back reaction of medium on magnetic field.

The paper is organized as follows. In  Section.~\ref{sec: weak B}, we derive the general  formulas of thermoelectric and electric conductivity tensors in an  (an-)isotropic medium at the  weak magnetic field  by solving the relativistic Boltzmann equation under the relaxation time approximation (RTA).
And the general expressions of the Seebeck coefficient and the  Nernst signal are presented. In Section.~\ref{sec:strong B},  using the same methodological in Section.~\ref{sec: weak B},  we also deduce the  formulas of both longitudinal tensors and  the longitudinal Seebeck coefficient, in the strong magnetic field with  the  Landau quantization.  In Section.~\ref{sec:tau}, the generalized expression of the  thermal relaxation times  related to quark   chemical  potential and anisotropic parameter at different magnetic field regimes are given.
In Section.~\ref{sec:discussions}, we  discuss the qualitative and quantitative features  of  thermoelectric transport coefficients.
In Section.~\ref{sec:summary}, we present a summary and provide an outlook for the future.  More detailed derivation of the  relaxation time in  zero  magnetic field and in the  LLL approximation  can be  found in Appendix A and Appendix B, respectively.

\section{ thermoelectric coefficients in an (an-)isotropic medium at  weak magnetic field}\label{sec: weak B}
It is sufficient to calculate the thermoelectric coefficients using the  kinetic theory approach.
At weak magnetic field within the hierarchy of scale $eB\ll g^2T^2$,  the phase space and the single particle energy are intact by  magnetic field through the  Landau quantization.  The  magnetic field enters through the cyclotron frequency of the charged particles as in classical picture.  Accordingly,
the dynamic evolution of a single particle  distribution function $f_a$   in the  uniform electric field $\mathbf{E}$ and magnetic field $\mathbf{B}$ can be determined by the  relativistic Boltzmann equation~\cite{bfeng}
\begin{eqnarray}\label{eq:boltzmann}
\frac{\partial f_{a}}{\partial t}+\mathbf{v}_{a}\cdot \bm{\nabla}
	f_{a}+e_{a}[\mathbf{E}+\mathbf{v}_{a}\times\mathbf{B}]\cdot\frac{\partial 
	f_{a}}{\partial 
	\mathbf{p}}=C[f_{a}].
\end{eqnarray}
where, $e_{a}=q_{a}e$,  $q_{a}$ and  $\mathbf{v}_{a}\equiv\frac{d\epsilon_{a}}{d\mathbf{}}=\mathbf{p}/\epsilon_{a}$ are fractional charged value and the
velocity for  particle species $a$, respectively. $\epsilon_{a}=\sqrt{\mathbf{p}^2+m_{a}^2} $ is the  energy of particle species  $a$, where  $m_{a}$ is current mass of   $a$-th species.
 In the weak magnetic field, we do not consider the trivial setup in which  the electric field is parallel to the direction of magnetic field, i.e.,  $\mathbf{E}\parallel\mathbf{B}$ because  there is no Lorentz force term to bend the   trajectory of a charged particle.
 We take~$\mathbf{E} =(E_x,E_y,0)$ and $\mathbf{B}=(0,0,B)$ in the system so that the Hall effect and the Nernst effect can exist.
The right side  of Eq.~\ref{eq:boltzmann}  is the collision  term or collision integral, which describes the rate of change of the single-particle distribution   induced by scatterings.
We assume the system is close to the local thermodynamic equilibrium,  and introduce  the  commonly used relaxation time approximation (RTA). In the RTA, the collision term can be expressed as 
\begin{equation}
C[f_a]\simeq - 
\frac{\delta f_{a}}{\tau_{a}}.
\end{equation}
Here,  $\tau_{a}$ is the relaxation time of
species $a$ which describes how fast the system reaches the equilibrium again. And $\delta f_{a}$ is  the  infinitesimal change in distribution function due to external disturbance,
\begin{equation}
\delta f_a=f_a-\bar{f}_a,
\end{equation}
with  $\bar{f}_{a}$ being the local equilibrium  distribution function of $a$-th species. In  an isotropic medium or  an anisotropic medium caused by initial spatial expansion,    $\bar{f}_{a}$ in the local rest frame can be expressed as~\cite{Romatschke:2003ms}
\begin{equation}\label{eq:fbara}
\bar{f}_a(\mathbf{p},\mu_a)=\left\{\begin{array}{l}
\bar{f}^{0}_{a}= \frac{1}{e^{(\sqrt{\mathbf{p}^2+m_a^2}-\mu_{a})\beta}\pm1}\ , \quad\text{ ($\xi=0$)} \ \\
\bar{f}_{a}^{\xi}=\frac{1}{e^{(\sqrt{\mathbf{p}^2+\xi(\mathbf{p}\cdot \mathbf{n})^2+m_a^2}-\mu_{a})\beta}\pm 1}\ , \quad\text{ ($\xi\neq0$) } \
\end{array}\right.
\end{equation}
where $\beta=1/T$  and $\mu_{a}$ denote inverse temperature and  chemical potential of species $a$, respectively. $\pm$ in Eq.~(\ref{eq:fbara}) corresponds to fermions and 
bosons, separately. The  anisotropic parameter $\xi$ in Eq.~(\ref{eq:fbara}) is defined as~\cite{Romatschke:2003ms}
\begin{equation}
\xi=\frac{\langle\mathbf{p}_\perp^2\rangle}{2\langle p_{\parallel}^2\rangle}-1,
\end{equation} where $p_{\parallel}=\mathbf{p}\cdot\mathbf{n}$ and $\mathbf{p}_{\perp}=\mathbf{p}-\mathbf{n}(\mathbf{p}\cdot\mathbf{n})$ are the momentum components which are parallel and perpendicular to the momentum anisotropy  direction ($\mathbf{n}$), respectively.
 $\mathbf{p}=(|\mathbf{p}|\sin\theta\cos\phi,|\mathbf{p}|\sin\theta\sin\phi,|\mathbf{p}|\cos\theta)$, where we  use a notation $|\mathbf{p}|\equiv p$ for convenience.  $\mathbf{n}=(\sin\alpha,0,\cos\alpha)$, where $\alpha$ is  the angle between $z$ direction and $\mathbf{n}$ direction. Accordingly, $(\mathbf{p}\cdot\mathbf{n})^2=p^2(\sin\theta\cos\phi\sin\alpha+\cos\theta\cos\alpha)^2=p^2c(\theta,\phi,\alpha)$.  In this work, we consider the anisotropy direction is along the beam direction, so $\alpha$ is fixed as $\pi/2$, $\mathbf{n}=(1,0,0)$.  
Note that  $\xi>0$ stands for a contraction  of distribution function along $\mathbf{n}$ direction  whereas $-1<\xi<0$ represents the  stretching of distribution function  along $\mathbf{n}$ direction. 
As mentioned in Section.~\ref{sec:intro}, for the anisotropic QGP induced by the initial spatial expansion, $\xi$ is always larger than zero.
 In  the weak $\xi$ limit ($|\xi|\ll1$), $\bar{f}_a^{\xi}$ 
 can be expanded in Taylor series to leading order term of $\xi$,
\begin{eqnarray}
\bar{f}^{\xi}_a
=\bar{f}_{a}^{0}-\frac{\xi\beta }{2\epsilon_{a}}\bar{f}_{a}^{0}(1-\bar{f}_{a}^{0})(\mathbf{p}\cdot\mathbf{n})^2.
\end{eqnarray}
Considering  the  distribution function  and  chemical potential  are time independent and space-time independent, respectively, Eq.~(\ref{eq:boltzmann}) can  be rewritten as
\begin{eqnarray}\label{eq:boltzmann2}
\left[\nu_{a}-e_{a}B(v_{x}\frac{\partial}{\partial 
	p_{y}}-v_{y}\frac{\partial}{\partial 
	p_{x}})\right]f_{a}^{}&=&\nu_{a} \bar{f}_{a}^{}-e_{a}E_x\frac{\partial }{\partial 
	p_{x}}\bar{f}_{a}^{}\nonumber\\
-e_{a}E_y\frac{\partial }{\partial 
	p_{y}}\bar{f}_{a}^{}-(\mathbf{v}_{a}\cdot
\bm{\nabla}\bar{f}_a),
\end{eqnarray}
where $\nu_{a}=1/\tau_{a}$ is the inverse relaxation time of $a$-th species.  We further assume  the solution of Eq.~(\ref{eq:boltzmann2}) in an anisotropic medium
satisfies the following linear form,
\begin{eqnarray}\label{eq:fa}
f_{a}^{\xi}=\bar{f}_{a}^{\xi}-\frac{1}{\nu_{a}}e\mathbf{E}\cdot\frac{\partial 
	\bar{f}_{a}^{\xi}}{\partial 
	\mathbf{p}}-\mathbf{\Xi}\cdot\frac{\partial 
	\bar{f}_{a}^{\xi}}{\partial \mathbf{p}}-\frac{1}{\nu_{a}}\mathbf{v}\cdot
	\bm{\nabla}\bar{f}_{a}^{\xi },
\end{eqnarray}
with $\mathbf{\Xi}$ being  an unknown quantity related to magnetic field.
Inserting Eq.~(\ref{eq:fa}) into Eq.~(\ref{eq:boltzmann2}) and assuming  no temperature gradient exists along $z$-axis, we obtain
\begin{eqnarray}\label{eq:eq1}
0=\nu_{a}F_a(\Xi_{x}v_x+\Xi_{y}v_y)+\frac{\omega_{c,a}^{}F_a}{\nu_{a}}(eE_xv_{y}-eE_yv_{x})\nonumber\\
+\frac{\omega_{c,a}^{}}{\nu_{a}}G_a(\nabla_{y}Tv_x-\nabla_{x}Tv_y)+\omega_{c,a}^{}F_a(\Xi_{x}v_y-\Xi_{y}v_x),
\end{eqnarray}
where $\omega_{c,a}^{}=e_{a}B/\epsilon_{a}$ is the cyclotron frequency of species $a$.  The expressions of  $G_a$ and $F_a$ in Eq.~(\ref{eq:eq1}) can read as 
\begin{widetext}
\begin{eqnarray}
F_a&=&\beta \bar{f}_a^0(1\pm\bar{f}_a^{0})(1+\xi c(\theta,\phi,\alpha))-\frac{\xi\beta^2p^2c(\theta,\phi,\alpha)}{2\epsilon_{a}}\bar{f}_a^{0}(1\pm\bar{f}_a^{0})(1-\bar{f}_a^{0}\pm\bar{f}_a^0+\frac{1}{\beta \epsilon_{a}}),\\
G_a&=&\beta^2(\epsilon_{a}-\mu_a)\bar{f}_a^{0}(1\pm\bar{f}_a^{0})-\frac{\xi\beta^3p^2 c(\theta,\phi,\alpha)}{2\epsilon_{a}}(\epsilon_{a}-\mu_a)\bar{f}_a^{0}(1\pm\bar{f}_a^{0})(1-\bar{f}_a^{0}\pm\bar{f}_a^{0}-\frac{1}{\beta(\epsilon_{a}-\mu_a)}),
\end{eqnarray}
\end{widetext}
 Comparing the coefficients of $v_{x}$ and $v_{y}$, one gets
\begin{equation}\label{eq:vx}
\nu_{a}F_a\Xi_{x}-\omega_{c,a}^{}\frac{eE_y}{\nu_{a}}F_a-\omega_{c,a}^{}F_a\Xi_{y}+\frac{\omega_{c,a}^{}}{\nu_{a}}G_a\nabla_{y}T=0,
\end{equation}
\begin{equation}\label{eq:vy}
\nu_{a}F_a\Xi_{y}+\omega_{c,a}^{}\frac{eE_x}{\nu_{a}}F_a+\omega_{c,a}^{}F_a\Xi_{x}-\frac{\omega_{c,a}^{}}{\nu_a}G_a\nabla_{x}T=0.\\
\end{equation}
Solving Eqs.~(\ref{eq:vx}) and (\ref{eq:vy}), we obtain
\begin{eqnarray}\label{eq:Xixx}
\Xi_{x}&=&-\frac{\omega_{c,a}^{2}eE_x}{\nu_a(\omega_{c,a}^{2}+\nu_{a}^{2})}+\frac{\omega_{c,a}eE_y}{\omega_{c,a}^{2}+\nu_{a}^{2}}-
\frac{\omega_{c,a}^{}G_a}{F_a(\omega_{c,a}^{}+\nu_{a}^{2})}\nabla_{y}T\nonumber\\
&&+\frac{\omega_{c,a}^{2}G_a}{\nu_{a}F_a(\omega_{c,a}^{2}+\nu_{a}^{2})}\nabla_{x}T,
\end{eqnarray}
\begin{eqnarray}\label{eq:Xixy}
\Xi_{y}&=&-\frac{\omega_{c,a}^{}eE_x}{\omega_{c,a}^{}+\nu_{a}^{2}}-\frac{\omega_{c,a}^2eE_y}{\nu_a(\omega_{c,a}^{2}+\nu_{a}^{2})}+\frac{\omega_{c,a}^{2}G_a}
{F_a\nu_{a}(\omega_{c,a}^{2}+\nu_{a}^{2})}\nabla_{y}T\nonumber\\
&&+\frac{\omega_{c,a}^{}G_a}
{F_a(\omega_{c,a}^{2}+\nu_{a}^{2})}
\nabla_{x}T.
\end{eqnarray}
Inserting Eqs.~(\ref{eq:Xixx})-(\ref{eq:Xixy}) to  Eq.~(\ref{eq:fa}), we  finally obtain the following perturbative term,
\begin{eqnarray}\label{eq:deltaf}
\delta f^{\xi}_a=f_{a}^{\xi}-\bar{f}_{a}^{\xi}&=&\frac{\omega_{c,a}F_a}{(\omega_{c,a}^{2}
	+\nu_{a}^{2})}(eE_yv_x-eE_xv_y)\nonumber\\
&&+\frac{\nu_{a}F_a}{\omega_{c,a}^{2}+\nu_{a}^{2}}(eE_xv_x+eE_yv_y)\nonumber\\
&&+\frac{\omega_{c,a}G_a}{\omega_{c,a}^{2}+\nu_{a}^{2}}(\nabla_{x}Tv_y-\nabla_{y}Tv_x)\nonumber\\
&&-\frac{\nu_{a}G_a}{\omega_{c,a}^{2}+\nu_{a}^{2}}(\nabla_{x}Tv_x+\nabla_{y}Tv_y).
\end{eqnarray}
In a conducting medium,  the charge carriers moving along  the direction of thermal  current generated by the   thermal gradient
accumulate on the cold side  and an electric field  can be generated. This electric field in turn induces an electric current in the opposite direction to the thermal current, consequently, a net  electric current may  exist in the medium.
When a magnetic field perpendicular to the  thermal gradient 
(assuming the  thermal gradient is along $x$ direction) is applied,  the charge carriers moving along  the direction of  thermal gradient and along  induced electric field   can be deflected,  a   net electric  current perpendicular to  both the thermal gradient and magnetic field    can also be generated.
Hence, in the linear response theory, the general formula of electric current density ($\mathbf{J}_{a}$) for species $a$ in response to  electric field ($\mathbf{E}$)  and temperature gradient ($\bm{\nabla} T$)  is given by~\cite{ electric current4}
\begin{eqnarray}\label{eq:current}
\mathbf{J}_{a}=e_a\int\frac{d^{3}\mathbf{p}}{(2\pi)^{3}}\mathbf{v}\delta 
f_a=\hat{\sigma}_a\cdot\mathbf{E}
+\hat{\alpha}_a(-\bm{\nabla} T),
\end{eqnarray}
where $\hat{\sigma}_a$ and $\hat{\alpha}_a$ are the electrical conductivity tensors and  the thermoelectric  conductivity   tensors for species $a$, respectively.
And the electric current density can further  decompose as~\cite{decomposition}
	\begin{eqnarray}\label{eq:jx}
J_{x,a}&=&\sigma_{xx,a}E_{x}+\sigma_{xy,a}E_{y}+\alpha_{xx,a}(-\nabla_{x}T)\nonumber\\
&&+\alpha_{xy,a}(-\nabla_{y}T),
\end{eqnarray}
	\begin{eqnarray}\label{eq:jy}
J_{y,a}&=&\sigma_{yy,a}E_{y}+\sigma_{yx,a}E_{x}+\alpha_{yy,a}(-\nabla_{y}T)\nonumber\\
&&+\alpha_{yx,a}(-\nabla_{x}T).
\end{eqnarray}
The first term in Eq.~(\ref{eq:jx}) (Eq.~(\ref{eq:jy})) is the electric current due to the electric field along $x$ ($y$)-axis induced by the  more accumulating carriers on the cold side of  medium, and the third term  in Eq.~(\ref{eq:jx}) (Eq.~(\ref{eq:jy})) is the thermal current due to the thermal  gradient in  $x$ ($y$)-axis.
The second term and fourth term in Eq.~(\ref{eq:jx}) (Eq.~(\ref{eq:jy})) are deuterogenic terms   due to  the deflection of  the first term and third term  in Eq.~(\ref{eq:jy}) (Eq.~(\ref{eq:jx})) by the magnetic field directed along $z$-axis.

In the steady state (i.e., putting $J_{x,a}=J_{y,a}=0 $), to avoid complicating the unambiguous determination of the Seebeck coefficient and  the Nernst signal, we assume the thermal gradient  purely along  $x$-axis, i.e. $\nabla_{x} T \neq0$, $\nabla_{y}T=0$ (isothermal condition). Using 
the Onsager reciprocity relation of  the thermoelectric and electric conductivity tensors in a magnetic field, $\sigma_{xx,a}(\alpha_{xx,a})=\sigma_{yy,a}(\alpha_{yy,a})$ and  $\sigma_{xy,a}(\alpha_{xy,a})=-\sigma_{yx,a}(-\alpha_{yx,a})$, we finally derive the expressions of the Seebeck coefficient ($S_{xx,a}$) and  the Nernst signal ($N_a$)   for $a$-th species  from Eqs.~(\ref{eq:jx}) and~(\ref{eq:jy}) 
\cite{CR Wang,decomposition,A Kundu}
\begin{equation}\label{eq:Sxxa}
S_{xx,a}=\frac{E_{x}}{\nabla_xT}\bigg|_{J_{x,a}=J_{y,a}=0}=\frac{\sigma_{xx,a}\alpha_{xx,a}+\sigma_{xy,a}\alpha_{xy,a}}{\sigma_{xx,a}^{2}+\sigma_{xy,a}^{2}},
\end{equation}
\begin{eqnarray}\label{eq:Sxya}
N_a
=\frac{E_{y}}{\nabla_xT}\bigg|_{J_{x,a}=J_{y,a}=0}=\frac{\sigma_{xy,a}\alpha_{xx,a}-\sigma_{xx,a}\alpha_{xy,a}}{\sigma_{xx,a}^{2}+\sigma_{xy,a}^{2}}.\\\nonumber
\end{eqnarray}
Inserting Eq.~(\ref{eq:deltaf}) to Eq.~(\ref{eq:current}) and using  Eqs.~(\ref{eq:jx}) and~(\ref{eq:jy}), by intergrating over $\theta$ and $\phi$, we  get the expressions of the  electrical conductivity  ($\sigma_{xx,a}$) and  the Hall conductivity ($\sigma_{xy,a}$) of $a$-th species in an anisotropic medium, which can be presented in a  matrix form
\begin{widetext}
\begin{eqnarray}\label{eq:sigma-xx-xy}
\left(\begin{array}{cc}
\sigma_{xx,a}^{}\\
\sigma_{xy,a}^{}
\end{array}\right)=&&d_a\frac{e^2q_{a}^2\beta}{6}\int\frac{dp}{(\pi^2)}\frac{p^4}{\epsilon_{a}^2}\frac{1}{\omega_{c,a}^2+(\nu_{a}^\xi)^2}\left(\begin{array}{cc}
\nu_{a}^\xi\\
\omega_{c,a}
\end{array}\right)\bar{f}^{0}_a(1\pm\bar{f}^{0}_a)(1+\frac{\xi}{3})\nonumber\\
&&-d_a\frac{e^2q_{a}^2\beta^2\xi}{36}\int\frac{dp}{\pi^2}\frac{p^6}{\epsilon_{a}^3}\frac{1}{\omega_{c,a}^2+(\nu_{a}^\xi)^2}\left(\begin{array}{cc}
\nu_{a}^\xi\\
\omega_{c,a}
\end{array}\right)f^0_a(1\pm\bar{f}^{0}_a)(1-\bar{f}^{0}_a\pm\bar{f}_a^{0}+\frac{1}{\beta\epsilon_{a}}),
\end{eqnarray}
\end{widetext}
where $d_{a}$  and $\nu_{a}^{\xi}$ are  degeneracy  factor and the $\xi$-dependent inverse relaxation time  for particle species $a$, respectively. When $\xi=0$, Eq.~(\ref{eq:sigma-xx-xy}) reduces to the  stardand form in the isotropic medium,
\begin{eqnarray}\label{eq:sigma-xx-xy-0}
\left(\begin{array}{cc}
\sigma_{xx,a}^{}\\
\sigma_{xy,a}^{}
\end{array}\right)&=&d_a\frac{e^2q_{a}^2\beta}{3}\int\frac{d^3\mathbf{p}}{(2\pi)^3}\frac{p^2}{\epsilon_{a}^2}\frac{1}{\omega_{c,a}^2+(\nu_{a})^2}\left(\begin{array}{cc}
\nu_{a}\\
\omega_{c,a}
\end{array}\right)\nonumber\\
&&\times\bar{f}^{0}_a(1\pm\bar{f}^{0}_a).
\end{eqnarray}
Accordingly,  the  thermoelectric conductivity ($\alpha_{xx,a}$)
and the Hall-like thermoelectric conductivity ($\alpha_{xy,a}$) of $a$-th species  within the effect of momentum anisotropy  also can read as   
\begin{widetext}
\begin{eqnarray}\label{eq:alpha-xx-xy}
\left(\begin{array}{cc}
\alpha_{xx,a}^{}\\
\alpha_{xy,a}^{}
\end{array}\right)=&&d_a\frac{eq_{a}\beta^2}{6}\int\frac{dp}{\pi^2}\frac{p^4}{\epsilon_{a}^2}\frac{\epsilon_{a}-\mu_{a}}{\omega_{c,a}^2+(\nu_a^\xi)^2}\left(\begin{array}{cc}
\nu_{a}^\xi\\
\omega_{c,a}
\end{array}\right)\bar{f}_a^{0}(1\pm\bar{f}_a^{0})\nonumber\\
&&-d_a\frac{eq_{a}\beta^3\xi}{36}\int\frac{dp}{\pi^2}\frac{p^6}{\epsilon_{a}^3}\frac{1}{\omega_{c,a}^2+(\nu_a^\xi)^2}\left(\begin{array}{cc}
\nu_{a}^\xi\\
\omega_{c,a}
\end{array}\right)(\epsilon_{a}-\mu_{a})\bar{f}_a^{0}(1\pm\bar{f}_a^{0})(1-\bar{f}_a^{0}\pm\bar{f}_a^{0}-\frac{1}{\beta(\epsilon_{a}-\mu_{a})}).
\end{eqnarray}
\end{widetext}
In the isotropic medium, the matrix of the  thermoelectric conductivity  tensors  can be simplified as
\begin{widetext}
\begin{eqnarray}\label{eq:alpha-xx-xy-0}
\left(\begin{array}{cc}
\alpha_{xx,a}^{}\\
\alpha_{xy,a}^{}
\end{array}\right)=&&d_a\frac{eq_{a}\beta^2}{3}\int\frac{d^3\mathbf{p}}{(2\pi)^3}\frac{p^2}{\epsilon_{a}^2}\frac{\epsilon_{a}}{\omega_{c,a}^2+\nu_a^2}\left(\begin{array}{cc}
\nu_{a}\\
\omega_{c,a}
\end{array}\right)\bar{f}_a^{0}(1\pm\bar{f}_a^{0})-d_a\frac{eq_{a}\beta^2}{3}\int\frac{d^3\mathbf{p}}{(2\pi)^3}\frac{p^2}{\epsilon_{a}^2}\frac{\mu_{a}}{\omega_{c,a}^2+\nu_a^2}\left(\begin{array}{cc}
\nu_{a}\\
\omega_{c,a}
\end{array}\right)\bar{f}_a^{0}(1\pm\bar{f}_a^{0}).\nonumber\\
\end{eqnarray} 
\end{widetext}
 Since the total Seebeck coefficient ($S_{xx}$) and the total Nernst signal ($N$) are the sum of contributions from  different  types of carriers weighted by the respective  electrical conductivities~\cite{shuben}, in the  QGP with three-flavor (anti-)quarks   ($f= u,~d,~s$),  the expressions of $S_{xx}$ and $N$ can be naturally written as,
\begin{eqnarray}\label{eq:sxx}
S_{xx}&=&\frac{\sum_{f}^{}(S_{xx,q_f}\sigma_{xx,q_f}+S_{xx,\bar{q}_f}\sigma_{xx,\bar{q}_f})}{\sum_{f}(\sigma_{xx,q_f}+\sigma_{xx,\bar{q}_f})},\nonumber\\
N&=&\frac{\sum_{f}^{}(N_{q_f}\sigma_{xx,q_f}+N_{\bar{q}_f}\sigma_{xx,\bar{q}_f})}{\sum_{f}^{}(\sigma_{xx,q_f}+\sigma_{xx,\bar{q}_f})}.
\end{eqnarray}
The denominator in Eq.~(\ref{eq:sxx}) denotes  the total electrical conductivity ($\sigma_{xx}$), accordingly other total conductivity tensors  from different flavors (anti)-quarks also can be given as
$\sigma_{xy}=\sum_f(\sigma_{xy,q_f}+\sigma_{xy,\bar{q}_f})$,~$\alpha_{xx}=\sum_f(\alpha_{xx,q_f}+\alpha_{xx,\bar{q}_f})$,~$\alpha_{xy}=\sum_f(\alpha_{xy,q_f}+\alpha_{xy,\bar{q}_f})$.
When the magnetic field is turned off, the   Nernst effect is absent and only  the  Seebeck effect exists. At the same time, the expression of $S_{xx,a}$ reduces to $S_{xx,a}=\frac{\alpha_{xx,a}}{\sigma_{xx,a}}$.
We note that the conductivity tensors are coupling-constant-dependent, however,  the Seebeck coefficient and Nernst signal  are unaffected  by different  coupling constants. It can be understood that the coupling constant term, which is  embedded in the relaxation time,   in the   numerator  of  Eq.~(\ref{eq:sxx})  is exactly cancelled by  that  in the denominator.

\section{thermoelectric coefficients in an (an-)isotropic medium at strong magnetic field}\label{sec:strong B}
 At strong magnetic field directed in $z$-axis  within the regime  $eB\gg T^2$, in the significant Landau quantization, if we still take  electric field along  $x$-direction, the electrical conductivity ($\sigma_{xx}$)  is zero  in  the one-loop  calculation~\cite{Harutyunyan:2016rxm} and the  Hall conductivity becomes $\sigma_{xy}=n_e/B$ ($n_e$ is electron number density)~\cite{electrical3}.
 Very recently, S.~Lin and L. Yang have deduced the general formulas of  $\sigma_{xx}$ and  $\sigma_{xy}$ in  chiral kinetic theory with full Landau level basis at the strong magnetic field along $z$-axis~\cite{Lin:2019fqo}. However, due to the lack of corresponding collision term,  it still remains a great challenge to fully understand the flavor dynamics of a magnetized QCD plasma in the background of strong magnetic field perpendicular to electric field.  Accordingly, we  focus on a simply setup in which the direction of  electric field  is  parallel to the direction of magnetic field.
Therefore, the linear Boltzmann equation of the magnetic-field-dependent single  particle distribution $f_{B,a}^{}$ at an external electric field $\mathbf{E}=(0,0,E_z)$ in  the RTA  can be given as
\begin{equation}
\frac{\partial f_{B,a}^{}}{\partial t}+v_{z}\nabla_z f_{B,a}+e_aE_{z}\frac{\partial f_{B,a}^{}}{\partial p_z}=-\frac{\delta f^{}_{B,a}}{\tau_{B,a}},
\end{equation}
 where the correction term $\delta f_{B,a}^{}$ can be written as 
\begin{eqnarray}\label{eq:deltaf'}
\delta f^{}_{B,a}=-\tau_{B,a}^{}[v_{z}\nabla_{z}\bar{f}^{}_{B,a}+eq_{a}E_{z}\frac{\partial \bar{f}^{}_{B,a}}{\partial p_z}].
\end{eqnarray}
 Due to the motion of (anti-)quarks in the Landau quantization is mainly restricted  to the direction of magnetic field,
 the  equilibrium   distribution function of charge particle  $\bar{f}_{B,a}^{}$ in an isotropic medium  and in  an anisotropic medium induced by strong magnetic field
can read as 
\begin{equation}\label{f-B}
\bar{f}_{B,a}(p_z,\mu_{a})=\left\{\begin{array}{l}
\bar{f}^{0}_{B,a}= \frac{1}{e^{(\epsilon_{a,l}-\mu_{a})\beta}\pm 1},  (\xi'=0)\ ,\\
\bar{f}_{B,a}^{\xi'}=\frac{1}{e^{(\sqrt{\epsilon_{a,l}^2+\xi'(\mathbf{p}\cdot \mathbf{n'})^2}-\mu_{a})\beta}\pm 1},(\xi'\neq0).
\end{array}\right.
\end{equation}
Here we assume that   the direction of momentum anisotropy $\mathbf{n}'$ is directed in the direction of magnetic field, and  $\mathbf{p}\approx(0,0,p_{z})$, as Ref.\cite{anisotropic-B}.
 Unlike the anisotropic parameter $\xi$,  the anisotropic parameter induced by strong magnetic field ($\xi'$) is always  negative because the momentum component along the direction of momentum anisotropy (viz, along the direction of magnetic field)  is dominant as compared to that  along other directions.
In the small $\xi'$ limit ($|\xi'|\ll1$), $\bar{f}_{B,a}^{\xi'}$ also can be  expanded  to leading order of $\xi'$,
\begin{equation}\label{eq:fxi'}
\bar{f}^{\xi'}_{B,a}=\bar{f}^{0}_{B,a}-\frac{\xi'\beta p_{z}^2}{2\epsilon_{a,l}}\bar{f}^{0}_{B,a}(1\pm\bar{f}^{0}_{B,a}).
\end{equation}

The phase space integration  for $a$-th charged particle   due to the   dimensional reduction of  motion in  Landau quantization is modified to
$\int\frac{d^3\mathbf{p}}{(2\pi)^3}=\sum_{l=0}^{\infty}\frac{|q_{a}eB|}{2\pi}\int_{-\infty}^{+\infty}\frac{dp_z}{2\pi}$, where $eB/(2\pi) $ is the density of states in two spatial directions perpendicular  to the direction of magnetic field~\cite{Andersen:2014xxa,Chakrabarty:1996te}. In the linear response theory, the longitudinal electric current  density of species $a$  can be written as
\begin{eqnarray}\label{eq:jz}
J_{z,a}&=&\sum_{l}g_{a,l}q_{a}e\frac{|q_{a}eB|}{2\pi}\int_{-\infty}^{+\infty}\frac{dp_z}{2\pi}v_{z}\delta f^{}_{B,a}\nonumber\\
&=&\sigma_{zz,a}E_{z}+\alpha_{zz,a}(-\nabla_{z}T),
\end{eqnarray}
 with  $g_{a,l}$ being the degeneracy factor of $a$-th species. For quarks and anti-quarks, $g_{a,l}=\sum_{f}(2-\delta_{0l}) N_{c}$, in which  $(2-\delta_{0l})$ and $N_{c}$ are spin degeneracy factor of  the Landau levels and the number of quark colors, repsectively. Since the direction of temperature gradient  is parallel to the direction of magnetic field in the strongly magnetized medium, the Nernst effect vanishes. Setting $J_{zz,a}=0$, the longitudinal  Seebeck coefficient of  $a$-th species, $S_{zz,a}$, can be expressed as $S_{zz,a}=\frac{E_z}{\nabla_zT}=\frac{\alpha_{zz,a}}{\sigma_{zz,a}}$, where $\sigma_{zz,a}$ ($\alpha_{zz,a}$) is the longitudinal  electrical (thermoelectric) conductivity of $a$-th species. 
 Using Eqs.~(\ref{eq:deltaf'})-(\ref{eq:jz})  the  formulas of $\sigma_{zz,a}$ and  $\alpha_{zz,a}$ in the (an-)isotropic medium can be respectively derived as,
\begin{widetext}
\begin{eqnarray}
\sigma_{zz,a}^{}=&&-^{}e^2q_{a}^2\frac{|q_aeB|\beta^2\xi'}{2\pi}\sum_{l}^{}g_{a,l}\int_{-\infty}^{+\infty}\frac{dp_{z}}{2\pi}\frac{p_{z}^4}{2\epsilon_{a,l}^3}\tau_{B,a}^{\xi'}\bar{f}_{B,a}^{0}(1\pm\bar{f}_{B,a}^{0})(1-2\bar{f}_{B,a}^{0}+\frac{1}{\beta\epsilon_{a,l}})\nonumber\\
&&+e^2q_{a}^2\frac{|q_aeB|\beta}{2\pi}\sum_{l}^{}g_{a,l}\int_{-\infty}^{+\infty}\frac{dp_{z}}{2\pi}
\frac{p_{z}^2}{\epsilon_{a,l}^2}\tau_{B,a}^{\xi'}\bar{f}_{a,B}^{0}(1\pm\bar{f}_{B,a}^{0})(1+\xi'),\quad (\mathrm{ for}~ \xi'\neq 0);\\
=&&e^2q_{a}^2\frac{|q_aeB|\beta}{2\pi}\sum_{l}^{}g_{a,l}\int_{-\infty}^{+\infty}\frac{dp_{z}}{2\pi}\frac{p_{z}^2}{\epsilon_{a,l}^2}\tau_{B,a}^{}\bar{f}_{B,a}^{0}(1\pm\bar{f}_{B,a}^{0}),\quad (\mathrm{ for}~ \xi'=0);
\end{eqnarray}
\end{widetext}
\begin{widetext}
\begin{eqnarray}
\alpha_{zz,a}^{}=&&-eq_{a}\frac{|q_aeB|\beta^3\xi'}{2\pi}\sum_{l}^{}g_{a,l}\int_{-\infty}^{+\infty}\frac{dp_{z}}{2\pi}\frac{p_{z}^4}{2\epsilon_{a,l}^3}\tau_{B,a}^{\xi'}(\epsilon_{a,l}-\mu_{a})\bar{f}_{B,a}^{0}(1\pm\bar{f}_a^{0})(1-2\bar{f}_{B,a}^{0}-\frac{1}{\beta(\epsilon_{a,l}-\mu_{a})})\nonumber\\
&&+eq_{a}\frac{|q_aeB|\beta^2}{2\pi}\sum_{l}^{}g_{a,l}\int_{-\infty}^{+\infty}\frac{dp_{z}}{2\pi}\frac{p_{z}^2}{\epsilon_{a,l}^2}\tau_{B,a}^{\xi'}(\epsilon_{a,l}-\mu_{a})\bar{f}_{B,a}^{0}(1\pm\bar{f}_{B,a}^{0}),
\quad (\mathrm{ for}~ \xi'\neq 0);\\
=&&eq_{a}\frac{|q_aeB|\beta^2}{2\pi}\sum_{l}^{}g_{a,l}\int_{-\infty}^{+\infty}\frac{dp_{z}}{2\pi}\frac{p_{z}^2}{\epsilon_{a,l}^2}\tau_{B,a}^{}(\epsilon_{a,l}-\mu_{a})\bar{f}_{B,a}^{0}(1\pm\bar{f}_{B,a}^{0}),\quad (\mathrm{ for}~ \xi'=0).
\end{eqnarray}
\end{widetext}
Here  $\tau_{B,a}(\tau_{B,a}^{\xi'})$ is the magnetic-field-dependent  relaxation time of species $a$  in the isotropic (anisotropic) medium.
Finally,  the total longitudinal Seebeck coefficient in  the QGP with three-flavor  can be given as
\begin{eqnarray}\label{eq:szz}
S_{zz}^{}&=&\frac{\sum_{f}(S_{zz,q_f}^{}\sigma_{zz,q_f}^{}+S_{zz,\bar{q}_f}^{}\sigma_{zz,\bar{q}_f}^{})}{\sum_{f}(\sigma_{zz,q_f}^{} +\sigma_{zz,\bar{q}_f}^{})  
}\nonumber\\
&&=\frac{\sum_{f}^{}(\alpha_{zz,q_f}^{}+\alpha_{zz,\bar{q}_f}^{})}{\sum_{f}^{}(\sigma_{zz,q_f}^{}+\sigma_{zz,\bar{q}_f })}=\frac{\alpha_{zz}}{\sigma_{zz}},
\end{eqnarray}
with $\alpha_{zz}$ ($\sigma_{zz}$) being the  total longitudinal thermoelectric (electrical) conductivity.

\section{thermal relaxation time }\label{sec:tau}
In this work,  the   relaxation time is a vital dynamic input for the calculation of thermoelectric coefficients in the QGP. The computation of relaxation time inevitably involves the choice of effective  running coupling constant, which can control the behavior of transport parameters critically. The effect of  momentum anisotropy also can enter the effective coupling constant through the   calculation of  Debye mass using the anisotropic distribution functions.
Conventionally,  the Debye screening mass is obtained by the static limit of the gluon self-energy in Hard Thermal Loop (HTL) theory~\cite{HTL}. In this work, we use a parallel approach, i.e., the  semi-classical transport theory~\cite{semi-transport}  to get it.  
Furthermore, as the Debye screening mass  manifests itself in the collective oscillation of the medium through the dispersion relation~\cite{Hasan:2017fmf},
	accordingly it can be affected by the Landau quantization.
As mentioned in Section.~\ref{sec:intro},  the scattering processes  are significantly different at weak and strong magnetic fields, therefore   we split this section into two separate parts: (A) without Landau quantization, and (B) with Landau quantization. 
The  effects of momentum anisotropy induced by initial spatial expansion and strong  magnetic field also straightly enter into the relaxation time and the Debye mass through  replacing  the isotropic distribution functions in associated expressions with  the momentum anisotropic counterparts.

\subsection{Without Landau quantization}
At  weak magnetic field  within the regime  $g^2T^2\gg eB$, the motions of particles in the
QGP  are not affected by the magnetic field through the Landau quantization. The  Debye mass ($m_D$)  in the  isotropic medium  for vanishing magnetic field can be expressed as~\cite{Kurian:2017yxj}
 \begin{eqnarray}\label{eq:MD0}
 m_{D}^2=-4\pi\alpha_{s}\int\frac{d^3\mathbf{p}}{(2\pi)^3}\left[2N_{c}\frac{d\bar{f}^0_{g}}{dp}
 +\sum_{f}\sum_{i=q_f,\bar{q}_f}\frac{d\bar{f}^0_{i}}{dp}\right],
 \end{eqnarray}
 where  $\mu_{q}$ ($\mu_{\bar{q}}$) is quark (anti-quark) chemical potential and $\mu_{\bar{q}}=-\mu_{q}$. 
 When $\mu_{q}/T\lesssim1$, $ m_{D}^2=4\pi\alpha_{s}T^2\left[(\frac{N_{c}}{3}+\frac{N_{f}}{6})+(\frac{\mu_{q}}{T})^2\frac{N_{f}}{2\pi^2}\right]$ for the massless case, where $N_{f}$ represents the number of quark flavors. This result is equal to the leading-order result in HTL approximation~\cite{HTL2,HTL3,HTL-1978}. 
 In the momentum anisotropic medium induced by initial spatial expansion,
 the Debye mass needs to be modified by roughly replacing the isotropic distribution function with the anisotropic distribution function, which  can be written as
 \begin{eqnarray}\label{eq:MD-xi}
 (m_{D}^{\xi})^2
 =4\pi\alpha_{s}\int\frac{d^3\mathbf{p}}{(2\pi)^3}[2N_cF_g+N_f(F_q+F_{\bar{q}})].
 \end{eqnarray}
In  an anisotropic medium induced by initial spatial expansion, the $\xi$-dependent effective   coupling constant  can be defined as $\alpha_{eff}^{}=\frac{(m_{D})^2}{(m_{D}^\xi)^2}\alpha_{s}$, where  $\alpha_{s}=\frac{6\pi}{(33-2N_{f})\ln\left(\frac{2\pi T}{\Lambda_{\bar{MS}}}\right)}$ is the one-loop running coupling at  $\Lambda_{\bar{MS}}=176~\mathrm{MeV}$ for $N_f=3$~\cite{Bazavov}. 
For the elastic  process  $a_1(P_1)+a_2(P_2)\rightarrow a_3(P_3)+a_4(P_4)$ ($P_{i=1,2,3,4}$ is the  four-momentum of $i$-th particle), the inverse relaxation time of species $a_1$, $1/\tau_{1}$,
 can be given as (see Appendix A for the detailed derivation)
  \begin{align}
\tau_{1}^{-1}=\sum_{pro}\frac{d_2}{\delta_{34}+1}\int\frac{d^3\mathbf{p}_2}{(2\pi)^3}\bar{f}_{2}(1\pm\bar{f}_4)\int dt \frac{d\sigma^{pro}}{dt}\frac{2tu}{s^2},
\end{align}
where $s,t,u$ are  Mandelstam variables,  and $\frac{d\sigma^{pro}}{dt}$ is the differential cross section for a specific scattering process. In our work   only  the   elastic scatterings, $viz$, 
  (1) $gq\rightarrow gq$, (2) $qq\rightarrow qq$,  (3) $q\bar{q}\rightarrow q\bar{q}$, (4)$qq'\rightarrow  qq'$, (5) $q\bar{q}'\rightarrow q\bar{q}'$ are considered.  We reasonably discard the inelastic processes like $q\bar{q}\rightarrow gg$ due to their small contributions. 
  The differential  cross sections in the leading order   perturbative QCD calculation for  the massless case  can be found in Ref.~\cite{pQCD}.
  Finally,  the momentum-averaged thermal relaxation time of (anti-)quarks  in the isotropic medium can be written as
  \begin{widetext}
\begin{eqnarray}\label{eq:tau-q}
\nu_{q(\bar{q})}=\tau_{q(\bar{q})}^{-1}&=&d_{g}\int\frac{d^3\mathbf{p}}{(2\pi)^3}\bar{f}_{g}^0(\mathbf{p})\left(1+\bar{f}_{g}^0(\mathbf{p})\right)\frac{1}{4}\frac{g^4}{\pi\langle s_{q(\bar{q})g}\rangle}\left[\ln\frac{\langle s_{q(\bar{q})g}\rangle}{\mu_{D}^2}-\frac{139}{108}\right]\nonumber\\
&&+d_{q(\bar{q})}\int\frac{d^3\mathbf{p}}{(2\pi)^3}\bar{f}^0_{q(\bar{q})}(\mathbf{p},\mu_{q(\bar{q})})\left(1-\bar{f}^0_{q(\bar{q})}(\mathbf{p},\mu_{q(\bar{q})})\right)\frac{1}{9}\frac{g^4}{\pi\langle s_{qq(\bar{q}\bar{q})}\rangle}\left[\ln\frac{\langle s_{qq(\bar{q}\bar{q})}\rangle}{\mu_{D}^2}-\frac{152}{96}\right]\nonumber\\
&&+d_{\bar{q}(q)}\int\frac{d^3\mathbf{p}}{(2\pi)^3}\bar{f}^0_{\bar{q}(q)}(\mathbf{p},\mu_{\bar{q}(q)})\left(1-\bar{f}^0_{\bar{q}(q)}(\mathbf{p},\mu_{\bar{q}(q)})\right)\frac{1}{9}\frac{g^4}{\pi\langle s_{q\bar{q}}\rangle}\left[\ln\frac{\langle s_{q\bar{q}}\rangle}{\mu_{D}^2}-\frac{147}{120}\right]\nonumber\\
&&+d_{q'(\bar{q'})}\int\frac{d^3\mathbf{p}}{(2\pi)^3}\bar{f}^0_{q'(\bar{q'})}(\mathbf{p},\mu_{q(\bar{q})})\left(1-\bar{f}^0_{q'(\bar{q'})}(\mathbf{p},\mu_{q(\bar{q})})\right)\frac{1}{9}\frac{g^4}{\pi\langle s_{qq(\bar{q}\bar{q})}\rangle}\left[\ln\frac{\langle s_{qq(\bar{q}\bar{q})}\rangle}{\mu_{D}^2}-\frac{17}{12}\right]\nonumber\\
&&+d_{\bar{q'}(q')}\int\frac{d^3\mathbf{p}}{(2\pi)^3}\bar{f}^0_{\bar{q'}(q')}(\mathbf{p},\mu_{\bar{q}(q)})\left(1-\bar{f}^0_{\bar{q'}(q')}(\mathbf{p},\mu_{\bar{q}(q)})\right)\frac{1}{9}\frac{g^4}{\pi\langle s_{q\bar{q}}\rangle}\left[\ln\frac{\langle s_{q\bar{q}}\rangle}{\mu_{D}^2}-\frac{17}{12}\right],
\end{eqnarray}
\end{widetext}
where   $\langle s_{ij}\rangle=2\langle p_{i}\rangle\langle p_{j}\rangle$ and $\langle p_i\rangle=\frac{\int\frac{d^3\mathbf{p}_i}{(2\pi)^3}|\mathbf{p}_i|
	\bar{f}_i^0}{\int\frac{d^3\mathbf{p}}{(2\pi)^3}\bar{f}^0_i}$ is the thermal average value of $p_i$.
And  in Eq.~(\ref{eq:tau-q})   the degeneracy factors of gluons and (anti-)quarks are $d_{g}=2_{spin}\times (N_{c}^2-1)$ and $d_{q(\bar{q})}=2_{spin}\times N_{c}\times N_{f}$, respectively.   $q'(\bar{q}')$ denotes incoming  (anti-)quark, which is  different to another incoming (anti-)quark in flavor type. 
$\mu_{D}^2=g^2T^2$ denotes infrared regulator.
In the anisotropic medium  induced by initial spatial expansion, the $\xi$-dependent inverse relaxtion time of (anti-)quark, $\nu_{q(\bar{q})}^\xi$,  can be obtained by substituting  $\bar{f}_i^0$ and  $\alpha_{s}$ in Eq.~(\ref{eq:tau-q}) with  $\bar{f}_i^{\xi}$ and $\alpha_{eff}$, respectively. Consequently, the thermal average of $p_g$ as well as  $p_{q(\bar{q})}$ with anisotropic momentum distribution  can be rewritten as
\begin{eqnarray}
\langle p\rangle^{\xi}_{i=g,q,\bar{q}}=T\frac{\mathrm{ Li}_4(-e^{-\beta\mu_{i}})}{\mathrm{ Li}_3(-e^{-\beta\mu_{i}})}\frac{4\xi-6}{\xi-2},
\end{eqnarray}
with $\mathrm{Li}_{n}(z)$ being the PolyLog function.

\subsection{With  Landau quantization}
In the presence of strong magnetic field, only fermionic part of the Debye mass is affected by   the  Landau quantization,  Eq.~(\ref{eq:MD0}) can be  modified as
\begin{eqnarray}\label{eq:MDB}
m_{D,B}^2
&=&-4\pi\alpha_{s,B}\bigg[\int\frac{d^3\mathbf{p}}{(2\pi)^3}2N_{c}\frac{d}{dp}
\bar{f}^0_{g}+\frac{1}{2}\sum_{f}\frac{|q_{f}eB|}{2\pi}\nonumber\\
&&\times\sum_{l=0}^{\infty}(2-\delta_{0l})\int_{-\infty}^{+\infty}\frac{dp_{z}}{2\pi}\sum_{i=q_f,\bar{q}_f}\frac{d\bar{f}^0_{B,i}}{dp}\bigg],
\end{eqnarray}
 where a  QCD factor as $\frac{1}{2}$ is considered. 
In the regime of $T^2\ll eB$  within the LLL approximation,  the   running coupling constant  mainly depends  on magnetic field, which is given as~\cite{Rath:2017fdv,Ferrer:2014qka}
\begin{eqnarray}
\alpha_{s,B}^{-1}(B)&=&\alpha_{s}^0(\mu_0)^{-1}+\frac{11N_c}{12\pi}\ln\left(\frac{\Lambda_{QCD}^2+M_B^2}{\mu_0^2}\right)\nonumber\\
&&+\frac{1}{3\pi}\sum_{f}\frac{|q_feB|}{\sigma},
\end{eqnarray}
where $\alpha_{s}^0(\mu_0)=\frac{12\pi}{11N_c\ln(\frac{\mu_0^2+M_B^2}{\Lambda_{V}^2})}$, $M_B=1$~GeV  and $\sigma=0.18~ \mathrm{GeV^{2}}$ are  infrared mass and the string tension, respectively. In Refs.~\cite{Ferrer:2014qka}, $\Lambda_{V}$ and $\mu_0$ are taken as 0.385  and 1.1 GeV, respectively. In  the  LLL approximation with the  hierarchy of scale $ eB\gg T^2$, Eq.~(\ref{eq:MDB}) for the massless case reduces to
\begin{eqnarray}\label{eq:MB-0}
m_{D,B}^2
&=&4\pi\alpha_{s,B}T^2\frac{N_{c}}{3}+4\pi\alpha_{s,B}\sum_{f}\frac{|q_{f}eB|}{4\pi^2}\sum_{l=0}^{\infty}(2-\delta_{0l})\nonumber\\
&&\times\bigg[\frac{1}{e^{(\sqrt{2l|q_{f}eB|}+\mu_{q})/T}+1}+\frac{1}{e^{(\sqrt{2l|q_{f}eB|}-\mu_{q})/T}+1}\bigg]\nonumber\\
&=&4\pi\alpha_{s,B}[T^2\frac{N_{c}}{3}+\sum_{f}\frac{|q_{f}eB|}{4\pi^2}]\nonumber\\
&\approx&\alpha_{s,B}\sum_{f}\frac{|q_{f}eB|}{\pi},
\end{eqnarray} 
 which is  consistent with the  one-loop calculation  in the presence of strong magnetic field~\cite{Bandyopadhyay:2017cle,Hattori:2016lqx,Singh:2017nfa}.
Replacing $\bar{f}_{B,i}^0$ with $\bar{f}_{B,i}^{\xi'}$ in Eq.~(\ref{eq:MDB}), we  can get the following  $\xi'$-dependent Debye mass $m_{D,B}^{\xi'}$ in an anisotropic medium induced by strong magnetic field
\begin{eqnarray}
(m_{D,B}^{\xi'})^2&=&4\pi\alpha_{s,B}\sum_{f}\frac{|q_{f}eB|}{4T\pi}\int_{-\infty}^{+\infty}\sum_{l=0}^{\infty}\frac{dp_z}{2\pi}(2-\delta_{0l})\nonumber\\
&&\times\sum_{i=q,\bar{q}}H_{i},
\end{eqnarray}
where
$H_i=\bar{f}_{B,i}^{0}(1-\bar{f}_{B,i}^{0})(1+\xi')-\frac{\xi'\beta p_z^2}{2\epsilon_{i,l}}\bar{f}_{B,i}^{0}(1-\bar{f}_{B,i}^{0})(1-2\bar{f}_{B,i}^{0}+\frac{1}{\beta\epsilon_{i,l}})$.
 The effective coupling constant associated with momentum anisotropy  induced by strong magnetic  field  is also defined  as $\alpha_{eff,B}^{}=(m_{D,B}^{\xi'})^2\alpha_{s,B}/m_{D,B}^2$.
In the strong magnetic field within the LLL approximation, except for the usual elastic $2\rightarrow2$ processes,   the quark-antiquark pair to gluon $2\rightarrow1$ process  and vice versa  are  also allowed, which is  kinetically forbidden for  weak or zero magnetic field case because two massive particle cannot become a massless particle.   In the  strong magnetic field, due to  the   spatial dimensional mixmatch   between (anti-)quarks and gluons,   the  transverse momentum component of  the  gluon  acts as  " the gluon mass", so that the gluon  can be generated by two massive particles and vice versa.  And in the hierarchy $\alpha_{s,B}eB\ll T^2\ll eB$, when  $m_q^2\gg \alpha_{s,B} eB$, the $2\rightarrow2$  processes are subleading compared to the $2\rightarrow1$ process because  the typical scale of collision rate for $2\rightarrow 2$ processes $\propto \alpha_{s,B}^2$ is parametrically smaller than the  typical scale of $2\rightarrow1$ process $\propto \alpha_{s,B}$ (more details see Ref.~\cite{Hattori:2016lqx}). However, when $m_q^2\ll \alpha_{s,B} eB$,  the quark-antiquark $t$-channel scattering process and  the $2\rightarrow1$ process are  same order of running constant  $\alpha_{s,B}$. Therefore,  in present work, the collision terms with respect to  two  scattering processes, $viz$, quark-antiquark to gluon $q(P)+\bar{q}(P')\rightarrow g (K)$ and quark-antiquark $t$-channel scattering  $q (P)+\bar{q}(P'')\rightarrow q(P')+\bar{q}(P''')$ are considered under the restrictive hierarchy of scales $m_{q}^2\ll\alpha_{s,B}eB\ll T^2\ll eB$.
 Following Ref.~\cite{Hattori:2016lqx}, we  generalize
 the computation of the relaxation time  of $f$-th (anti-)quarks for $2\rightarrow1$ process  to the case of finite chemical potential ($\mu_{q}\neq0$) and anisotropic medium ($\xi'\neq0$), 
\begin{eqnarray}\label{eq:tau-21-LLL}
&\frac{1}{\tau_{B,q(\bar{q})}^{\xi'}}&\bigg|_{2\rightarrow1}(T,p_z,\mu_{q},l=0)\nonumber\\
&=&\frac{\alpha_{eff,B}^{}C_R m_{f}^{2}}{\epsilon_{f,0}\left(1-\bar{f}_{B,q(\bar{q})}^{\xi'}(p_z,\mu_{q(\bar{q})})\right)}\nonumber\\
&&\times\int_{-\infty}^{+\infty}\frac{dp'_z}{\epsilon_{f,0}'}\bar{f}_{B,\bar{q}(q)}^{\xi'}(p'_z,\mu_{\bar{q}(q)})\left(1+\bar{f}_{g}^{}(k)\right).\nonumber\\
\end{eqnarray}
Here $C_{R}=\frac{N_{c}^2-1}{2N_{c}}$ is the Casimir factor.  $f_g(k)=\frac{1}{e^{k/T}-1}$ with  $k=|\mathbf{k}|=\epsilon_{f,0}$+$\epsilon'_{f,0}$, where $\epsilon_{f,0}=\sqrt{p_z^2+m_f^2}$ and $\epsilon'_{f,0}=\sqrt{p_z'^2+m_f^2}$.  Furthermore, at nonzero $\mu_{q}$, the relaxation time  of  $f$-th (anti-)quarks   for 2~$\rightarrow$~2 process in an anisotropic medium   is given as (details in Appendix B)
\begin{eqnarray}\label{eq:tau-22-LLL}
&\frac{1}{\tau_{B,q(\bar{q})}^{\xi'}}&\bigg|_{2\rightarrow2}(T,p_z,\mu_{q},l=0)\nonumber\\
&&=\alpha_{eff,B}\frac{m_{f}^2}{\epsilon_{f,0}}\bigg(1-\bar{f}^{\xi'}_{B,\bar{q}(q)}(p_z,\mu_{\bar{q}(q)})\bigg).
\end{eqnarray}
We can  observe that  the inverse relaxation times  for $2\rightarrow1$  and  $2\rightarrow2$ processes are  of the same order of $\alpha_{eff,B}^{}$. 
Currently, the relaxation time  of (anti-)quarks for $2\rightarrow1$ process beyond  the LLL approximation at zero chemical potential has been  studied in a more realistic regime  $eB\gg g^2T^2$~\cite{electrical3,transport3,electrical5}. Similarly, we  extend  it to the case of $\mu_q\neq0$ and $\xi'\neq0$,
\begin{eqnarray}\label{eq:tau-nll}
&\frac{1}{\tau_{B,q(\bar{q})}^{\xi'}}&\bigg|_{2\rightarrow1}(T,p_z,\mu_{q},l)\nonumber\\
&=&\frac{1}{4\epsilon_{f,l}\left(1-\bar{f}_{B,q(\bar{q})}^{\xi'}(p_z,\mu_{q(\bar{q})},l)\right)}\nonumber\\
&&\times\sum_{l'\geq l}^{\infty}\int_{-\infty}^{+\infty}\frac{dp'_z}{2\pi}\frac{1}{2\epsilon'_{f,l'}}\bar{f}_{B,\bar{q}(q)}^{\xi'}(p'_z,\mu_{\bar{q}(q)},l')\nonumber\\
&&\times\bigg(1+\bar{f}_{g}^{\xi'}(k)\bigg)
X(l,l',H).
\end{eqnarray}
Here $H$  is defined as $H=\frac{(\epsilon_{q,l}+\epsilon_{\bar{q},l'})^2-(p_{z}+p_z')^2}{2e|q_{f}B|}$~\cite{electrical3,transport3}
 and $X(l,l',H)$ can read as
\begin{eqnarray}
X(l,l',H)&=&4\pi\alpha_{eff,B}^{}C_R\frac{l!}{l'!}e^{-H}H^{l'-l}\bigg(\bigg[4m_{f}^2\nonumber\\
&&-4|q_{f}eB|(l+l'-H)\frac{1}{H}(l+l')\bigg]F(l,l',H)\nonumber\\
&&+16|q_{f}eB|l'(l+l')\frac{1}{H}L_{l}^{l'-l}(H)L_{l-1}^{l'-l}(H)\bigg).  \nonumber\\
 \end{eqnarray}
For $l>0$, $F(l,l',H)=[L_{l}^{l'-l}(H)]^2+\frac{l'}{l}[L_{l-1}^{l'-l}(H)]^2$ as well as  $F(l,l',H)=1$ for $l=0$, where the function $L_{n}^{\alpha}(x)$ stands for the generalized Laguerre polynomial. In the limit of $eB\gg T^2$, $e^{-K}\approx1$, the  result  in Eq.~(\ref{eq:tau-nll}) is  consistent with   that in Eq.~(\ref{eq:tau-21-LLL}).
In this work, we take  thermal averaged relaxation time  $\langle\tau_{B,q}\rangle$ of quarks as dynamic input for the computation of  thermoelectric coefficients, which is defined as
\begin{equation}\label{eq:thermal-tau}
\langle\tau_{B,q}\rangle(T,\mu_{q})=\frac{1}{\langle\tau_{B,q}^{-1}\rangle}=\frac{\sum_{l=0}^{\infty}\int_{-\infty}^{+\infty} dp_z \bar{f}^{0}_{B,q(\bar{q})}(p_z)}{\sum_{l=0}^{\infty}\int_{-\infty}^{+\infty} dp_z\tau_{B,q}^{-1}(p_z) \bar{f}^{0}_{B,q}(p_z)}.
\end{equation}
In the anisotropic medium, the $\xi'$-dependent thermal averaged relaxation time of quark, $\langle\tau_{B,q}^{\xi'}\rangle$, also can be obtained by substituting $\bar{f}^{0}_{B,q}$ and  $\tau_{B,q}^{-1}$ in Eq.~(\ref{eq:thermal-tau}) with $\bar{f}^{\xi'}_{B,q}$ and  $(\tau_{B,q}^{\xi'})^{-1}$, respectively. And
in  the numerical calculation we artificially truncate the sum of Landau levels  at a finite maximum $l$.

\section{numerical result and discussion}\label{sec:discussions}
       \begin{figure*}
	\includegraphics[width=3in,height=2.5in]{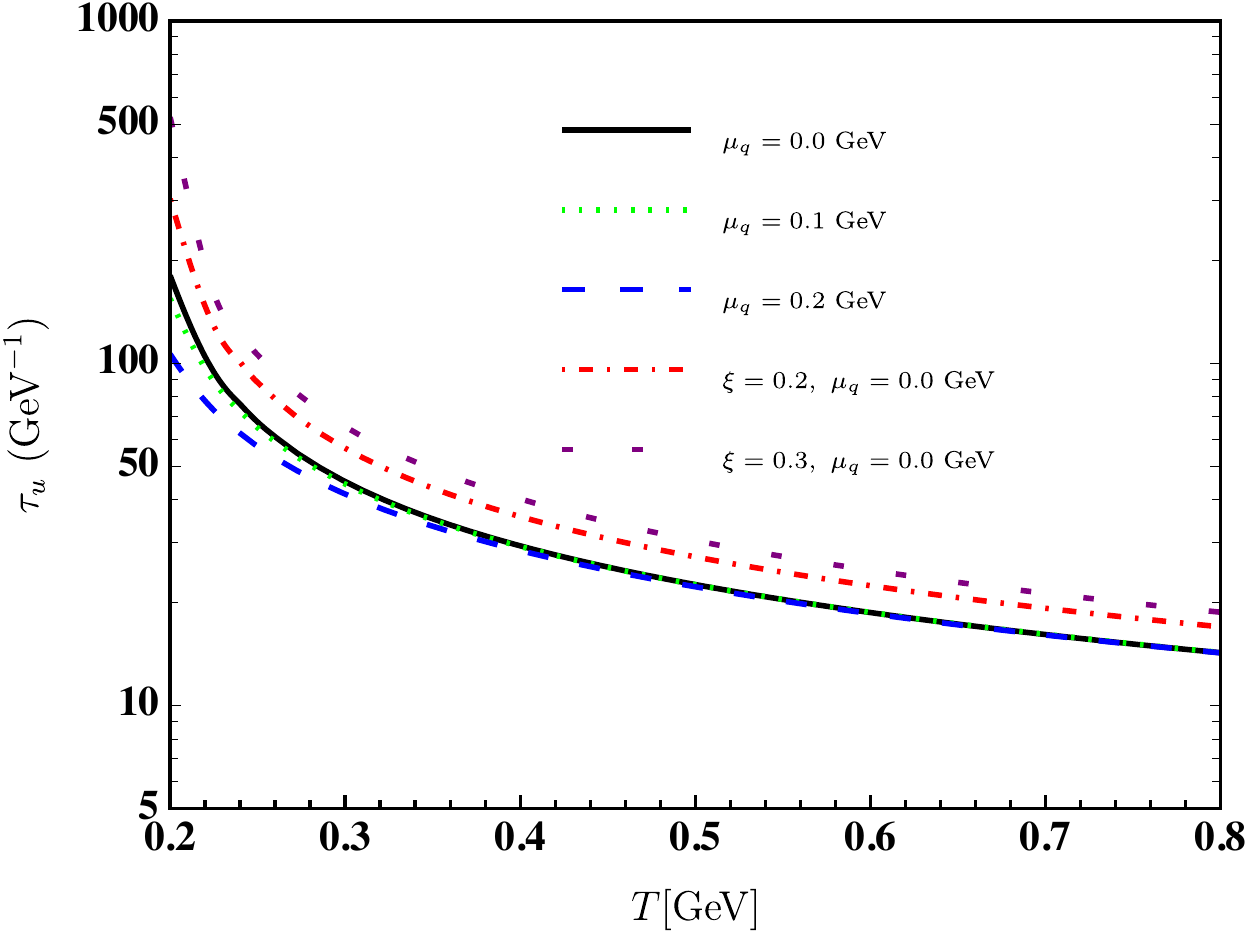}
	\caption{\label{fig:tau} The temperature dependence of the  thermal  relaxation time for $u$ quarks ($\tau_{u}$) at different quark chemical potentials, viz, $\mu_{q}$=0.0 GeV (black solid line), 0.1~GeV (green dotted line) and 0.2~GeV (blue dashed line) for $B=0$. The estimation of $\tau_{u}$ at $\mu_{q}=0.0 $~GeV is also extended to the anisotropic medium induced by initial spatial expansion with $\xi=$~0.2 (red dotted-dashed line) and 0.3 (purple short-dashed line).}
\end{figure*}
\begin{figure*}
	\includegraphics[width=7in,height=3.7in]{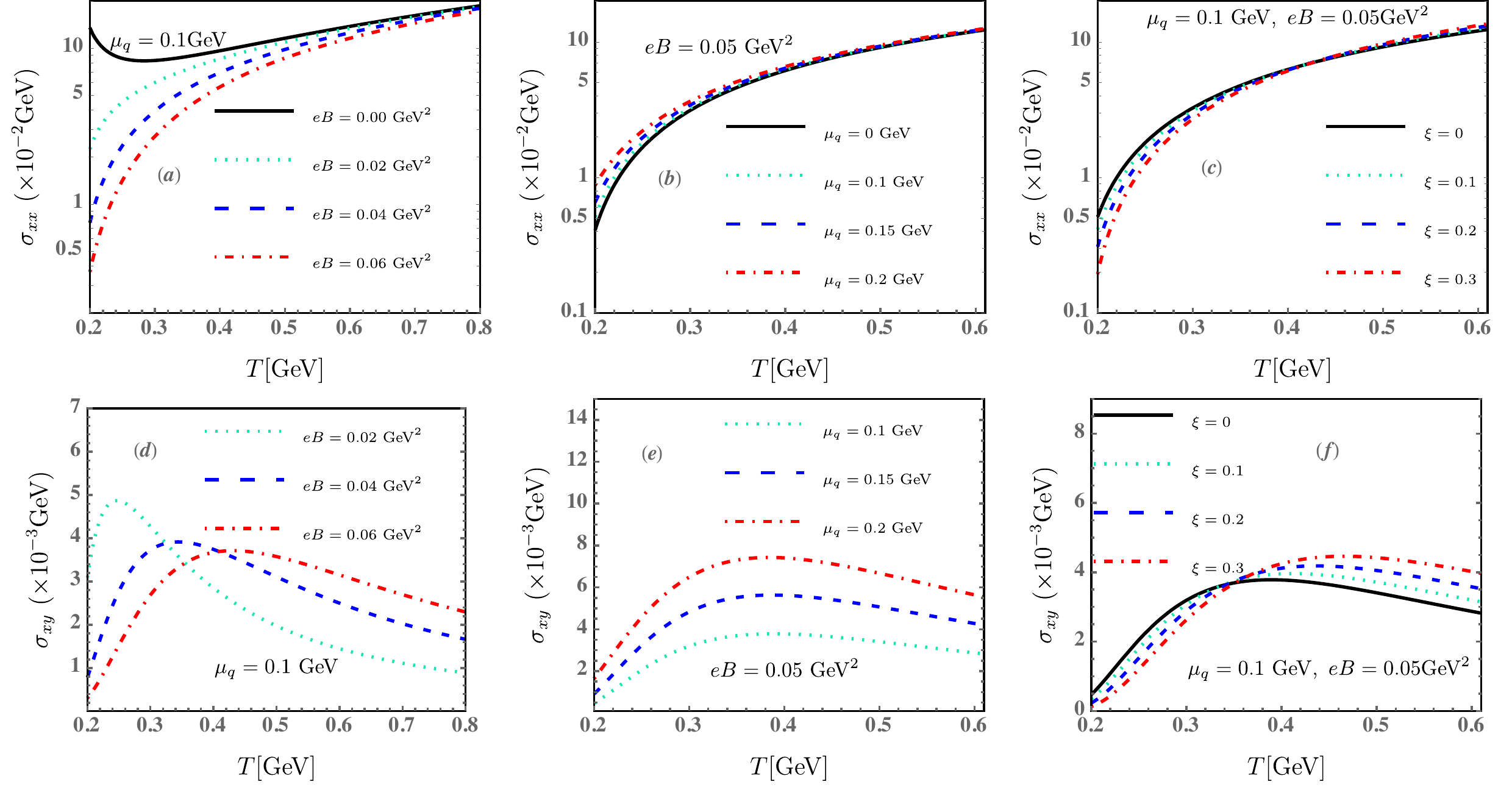}
	\caption{\label{fig:sigma}   [Diagrams (a) and (d)] The temperature dependences of the  electrical conductivity  ($\sigma_{xx}$) and  the Hall conductivity ($\sigma_{xy}$)  for $\mu_{q}=0.1$~GeV at different weak magnetic fields, namely, $eB=$ 0~$\mathrm{GeV^{2}}$ (black solid line), 0.02~$\mathrm{GeV^{2}}$ (cyan dotted lines), 0.04~$\mathrm{GeV^{2}}$ (blue dashed lines), 0.06~$\mathrm{GeV^{2}}$ (red dotted-dashed lines). [Diagrams  (b) and (e)] The temperature dependences of $\sigma_{xx}$  and $\sigma_{xy}$ for $eB=0.05~\mathrm{GeV^{2}}$ at  $\mu_{q}=$ 0~GeV (black solid line), 0.1~GeV (cyan dotted lines), 0.15~GeV (blue dashed lines) and 0.2~GeV (red dotted-dashed lines).  [Diagrams (c) and (f)] The temperature dependences of $\sigma_{xx}$ and $\sigma_{xy}$ for $\mu_{q}=0.1~$GeV and $eB=0.05~\mathrm{GeV^2}$ in a weakly anisotropic medium with $\xi=0$ (black solid lines), 0.1 (cyan dotted lines), 0.2 (blue dashed lines), 0.3 (red dotted-dashed lines).}
\end{figure*}
In the  numerical  calculation, we use the current masses  of three-flavor quarks   ($m_{u}=3~\mathrm{MeV}$,
$m_{d}=5~\mathrm{MeV}$ and $m_{s}=80~\mathrm{MeV}$) as input parameters.  As mentioned in Section.~\ref{sec:intro}, under  the weak magnetic field ($g^2T^2> eB$) all scatterings of partons in the   QGP   are  unaffected by  magnetic field, thus the  calculation of the relaxation time  remains the same as in the absence of the magnetic field. In Fig.~\ref{fig:tau}, the  thermal behavior of the relaxation time for $u$-quarks ($\tau_{u}$) at various quark chemical potentials ($\mu_{q}$) and anisotropic parameters ($\xi$) is displayed.  
 We observe that  $\tau_{u}$ decreases with increasing temperature and the order of magnitude of $\tau_{u}$ at small temperature is much larger than that at high temperature. We also notice $\tau_{u}$  decreases as $\mu_{q}$ increases at small temperature, whereas the decreasing feature of $\tau_{u}$ with $\mu_{q}$ is marginal at   high temperature  ($T>0.4~\mathrm{GeV}$).  This is because that with increasing temperature the ratio $\mu_q/T$ becomes smaller,  which leads to  the  result of different factors associated with  $e^{\pm\mu_{q}/T}$ in Eq.~(\ref{eq:tau-q}) is nearly $\mu_{q}$-independent. From Fig.~\ref{fig:tau} we  also can clearly see that  $\tau_{u}$ in an anisotropic QGP ($\xi=0.1$) has an overall improvement compared to  that  in an isotropic QGP, and the degree of improvement can be further strengthened with the increase of $\xi$. 
  To better understand the qualitative and quantitative behaviors of  the  total Seebeck coefficient ($S_{xx}$) and  the  total  Nernst signal ($N$) for weak magnetic field,  we first present our   results of  the  total electrical ($\sigma_{xx}$, $\sigma_{xy}$) and thermoelectric ($\alpha_{xx}$, $\alpha_{xy}$) conductivity tensors. 
  In Fig.~\ref{fig:sigma} (a), we perform the temperature  dependence of the total electrical conductivity ($\sigma_{xx}$)    at $\mu_{q}=0.1~$GeV for different weak magnetic fields.
   At the vanishing magnetic field, the thermal evolution of  $\sigma_{xx,{q_f}}$ for $f$-th flavor is basically dominated by the multiplicative result of factors, i.e., the relaxation time ($\tau_{q_f}$) and quark  distribution function ($\bar{f}^0_{q_f}$) in the integrand of  Eq.~(\ref{eq:sigma-xx-xy-0}). At small $T$, the sharply decreasing feature of $\tau_{q_f}(T)$  significantly wins over the increasing feature of $\bar{f}^0_{q_f}(T)$, therefore the total electrical conductivity ($\sigma_{xx}$) decreases with increasing temperature. However, at high $T$, the increasing behavior of $\bar{f}^0_{q_f}(T)$ is more  prominent than the  decreasing behavior  of $\tau_{q_f}(T)$.  As a result,   $\sigma_{xx}$  for vanishing magnetic field decreases at small $T$ then increases at high $T$, as shown in Fig.~\ref{fig:sigma} (a).
   In the presence of  $B$, we observe that   $\sigma_{xx}$ has a  suppression compared to that in the vanishing $B$. This is due to that a additional  factor,  $1/(1+(\omega_{c,q_f}\tau_{q_f})^2)$, 
   in integrand of Eq.~(\ref{eq:sigma-xx-xy-0}) is always less than 1.
    We see that  $\sigma_{xx}$ monotonously increases with increasing $T$ at nonzero $B$, which is different with the thermal behavior of $\sigma_{xx}$  at zero field. This  can be understood as follows: The qualitative behavior of $\sigma_{xx,{q_f}}$ for $f$-th flavor at weak $B$  is basically coming from the interplay between $\tau_{q_f}/(1+(\omega_{c,q_f}\tau_{q_f})^2)$ and $\bar{f}^0_{q_f}$.  At small $T$, $\tau_{q_f} $  is large and  $\sigma_{xx}(T)\sim\sum_{f}\tau_{q_f}\bar{f}^0_{q_f}/(1+(\omega_{c,q_f}\tau_{q_f})^2)\sim\sum_{f}\bar{f}^0_{q_f}/\tau_{q_f}$, consequently,  $\sigma_{xx}$ increases with increasing $T$. At high $T$, $\tau_{q_f} $ is relatively small, $\tau_{q_f}/(1+(\omega_{c,q_f}\tau_{q_f})^2)\sim\tau_{q_f} $, the thermal behavior of  $\sigma_{xx}$ at weak field is consistent with the counterpart  at zero field. Alternatively, the  dependence of $\sigma_{xx}$ on  $B$ only arises from the cyclotron frequency ($\omega_{c,q_f}$). Therefore,   $\sigma_{xx}$ decreases as $B$ grows at small $T$  due to  $\sigma_{xx}(B)\sim \sum_{f}\frac{1}{\omega_{c,q_f}^2}$, and the positive effect of $B$ on $\sigma_{xx}$ is unconspicuous at high $T$ due to $\sigma_{xx}(B)\sim$ constant. 
    Next, we consider the effect of  quark chemical potential ($\mu_{q}$) on the estimations of  the conductivity tensors at $eB=0.05~\mathrm{GeV^{2}}$. 
     We remind the reader that at  finite $\mu_{q}$ due to the  number density of quarks is  always larger than that of  anti-quarks in the QGP, the contribution of quarks to the  tensors in magnitude is prominent.
     At relatively small temperature, $\sigma_{xx}(\mu_{q})\sim\sum_{f}\bar{f}^0_{q_f}(\mu_{q})/\tau_{q_f}(\mu_{q})$, where  both $\bar{f}^0_{q_f}(\mu_q)$ and $1/\tau_{q_f}(\mu_{q})$ are   increasing  functions. Whereas, at high $T$, $\tau_{q_f}$ is  nearly unchanged with the variation of $\mu_{q}$ as shown in Fig.~\ref{fig:tau}, thus $\sigma_{xx}(\mu_{q})\sim\sum_{f}\bar{f}^0_{q_f}(\mu_{q})$. As the ratio $\mu_{q}/T$  at high $T$ is  small,  the  Boltzmann factor $e^{\mu_q/T}$  in quark  distribution  increases insignificantly  with an increase of $\mu_{q}$.  As a result, with the increase of $\mu_{q}$, $\sigma_{xx}$ increases  at  $T<0.3~\mathrm{GeV}$ then  remains almost constant  at $T>0.3~\mathrm{GeV}$, as shown in Fig.~\ref{fig:sigma} (b). 
    The  momentum  anisotropy  induced by initial spatial expansion is also  considered in the estimation of  the tensors.  In Fig.~\ref{fig:sigma} (c), the $T$ dependence of  $\sigma_{xx}$ at $\mu_q=0.1~\mathrm{GeV}$ and $eB=0.05$~$\mathrm{GeV^2}$ in the weakly anisotropic QGP (we take $\xi=0.1$, 0.2 and 0.3) is performed.
   As illustrated in Fig~\ref{fig:sigma} (c),  the  thermal behavior of $\sigma_{xx}$ in an anisotropic medium is consistent with the counterpart in an isotropic medium.   However,   the dependence of $\sigma_{xx}$ on $\xi$ is nonmonotonic in the entire $T$ domain of interest, which can be easily understood  from the expression of $\sigma_{xx,q_f}$ in Eq.~(\ref{eq:sigma-xx-xy}). Since  the first term in  Eq.~(\ref{eq:sigma-xx-xy}) is numerically much larger than the second term, thus the $\xi$ dependence  of $\sigma_{xx,{q_f}}$ is mainly determined by $\frac{\tau_{q_f}(\xi) }{1+\omega_{c,q_f}^2\tau_{q_f}(\xi)^2}(1+\xi)$ in the first term. At small $T$, $\sigma_{xx}(\xi)\sim\sum_{f}(1+\xi)/\tau_{q_f}(\xi)$, where  the increasing feature of (1+$\xi$) is compensated by the decreasing feature of $1/\tau_{q_f}(\xi)$, leading  $\sigma_{xx}$ as a decreasing function  of $\xi$. At high $T$, $\sigma_{xx}(\xi)\sim\sum_{f}\tau_{q_f}(\xi)(1+\xi)$, $\sigma_{xx}$  increases as  $\xi$ grows.

    Due to the absence of  the  Hall effect at the vanishing magnetic  field,  the calculation of the  total  Hall conductivity ($\sigma_{xy}$)  is only performed in the  magnetic background field. In the isotropic QGP, for the  Hall conductivity of  $f$-th flavor quarks, $\sigma_{xy,q_f}$, its thermal behavior   is mainly dominated by   the form factor  $\omega_{c,q_f}\tau_{q_f}^2/(1+(\omega_{c,q_f}\tau_{q_f} )^{2})$ and associated quark distribution ($\bar{f}^0_{q_f}$) in Eq.~\ref{eq:sigma-xx-xy-0}.  The numerator of the form factor  reminds us that the sign of   $\sigma_{xy,q_f}$ is dependent of quark type.
     Due to the discrepancies of various flavor quarks in fractional charge value ($q_{u,d,s}=2/3,-1/3,-1/3$) and mass, $\sigma_{xy,u}$ is greater than $\sigma_{xy,d}+\sigma_{xy,s}$.  This is why the sign of   $\sigma_{xy}$ always remains positive.
    In Fig.~(\ref{fig:sigma})(d)  we see that  $\sigma_{xy}$ at a fixed $B$ for  $\mu_{q}=0.1$~GeV exhibits a nonmonotonic thermal behavior. More exact, $\sigma_{xy}(T)$ first increases, reaches a maximum  then decreases. 
     This  peak structure of $\sigma_{xy}(T)$  has also been  observed in Ref.~\cite{bfeng}. We can understand this behavior in the following way.   At small $T$,  $\sigma_{xy}(T)\sim\sum_{f} \bar{f}^0_{q_f}$, where $\bar{f}^0_{q_f}$ is an increasing function of $T$. At high $T$, $\sigma_{xy}(T)\sim\sum_{f}\tau_{q_f}^2\bar{f}^0_{q_f}$ and the decreasing feature of $\tau_{q_f}^2(T)$  greatly overwhelms the increasing feature of $\bar{f}^0_{q_f}(T)$,  leading  $\sigma_{xy}$ as a decreasing function of $T$.
     From Fig.~\ref{fig:sigma} (d)  we also see that  as $B$ rises, $\sigma_{xy}$  decreases at small $T$ due to $\sigma_{xy}(B)\sim 1/B$, and increases  at high $T$ due to $\sigma_{xy}(B)\sim B$.  Moreover,  the position for the maximum  of $\sigma_{xy}$ shifts toward  higher $T$ with the increase in $B$.
      In  Fig.~\ref{fig:sigma} (e), $\sigma_{xy}$ increases as $\mu_{q}$ grows, which can be well understood from the behavior of  prominent $\sigma_{xy,u}(\mu_{q})$.
      In the  entire $T$ domain of interest, the qualitative behavior  of $\sigma_{xy,u}(\mu_{q})$  is  almost determined by  quark distribution function.
    Compared to $\sigma_{xx}(\mu_q)$,  we note that the effect of $\mu_{q}$ on $\sigma_{xy}$ is obvious at high temperature, which is attributed to the  increment in quark distribution function with the increase of  $\mu_{q}$ is comparable to the value of $\sigma_{xy}$ itself.
      In the momentum anisotropic medium,  the absolute value of $\sigma_{xy,q_f}$ for various flavors increases monotonously with the increase of $\xi$ because $\frac{|\omega_{c,q_f}|\tau_{q_f}(\xi)^2(1+\xi)}{1+(\omega_{c,q_f}\tau_{q_f}(\xi))^2}$ in Eq.~(\ref{eq:sigma-xx-xy})  is an increasing function of $\xi$.
      Although  $\sigma_{xy,u}$  itself is relatively larger than  $\sigma_{xy,d}+\sigma_{xy,s}$  in magnitude, the variation in $\sigma_{xy,d}+\sigma_{xy,s}$ with $\xi$ is numerically stronger (weaker) than the variation in  $\sigma_{xy,u}$ with $\xi$ at $T<0.3~\mathrm{GeV}$ (at $T>0.3~\mathrm{GeV}$)  (we don't display the figure, but it's a truth).  Hence, with the increase of $\xi$,   $\sigma_{xy}$ first decreases  at relatively low $T$ then increases  at high $T$, as shown in Fig.~\ref{fig:sigma} (f). We also observe that as $\xi$ grows, the maximum of  $\sigma_{xy}$ increases  and shifts  towards higher temperature.

      \begin{figure*}
      	\includegraphics[width=7.in,height=3.7in]{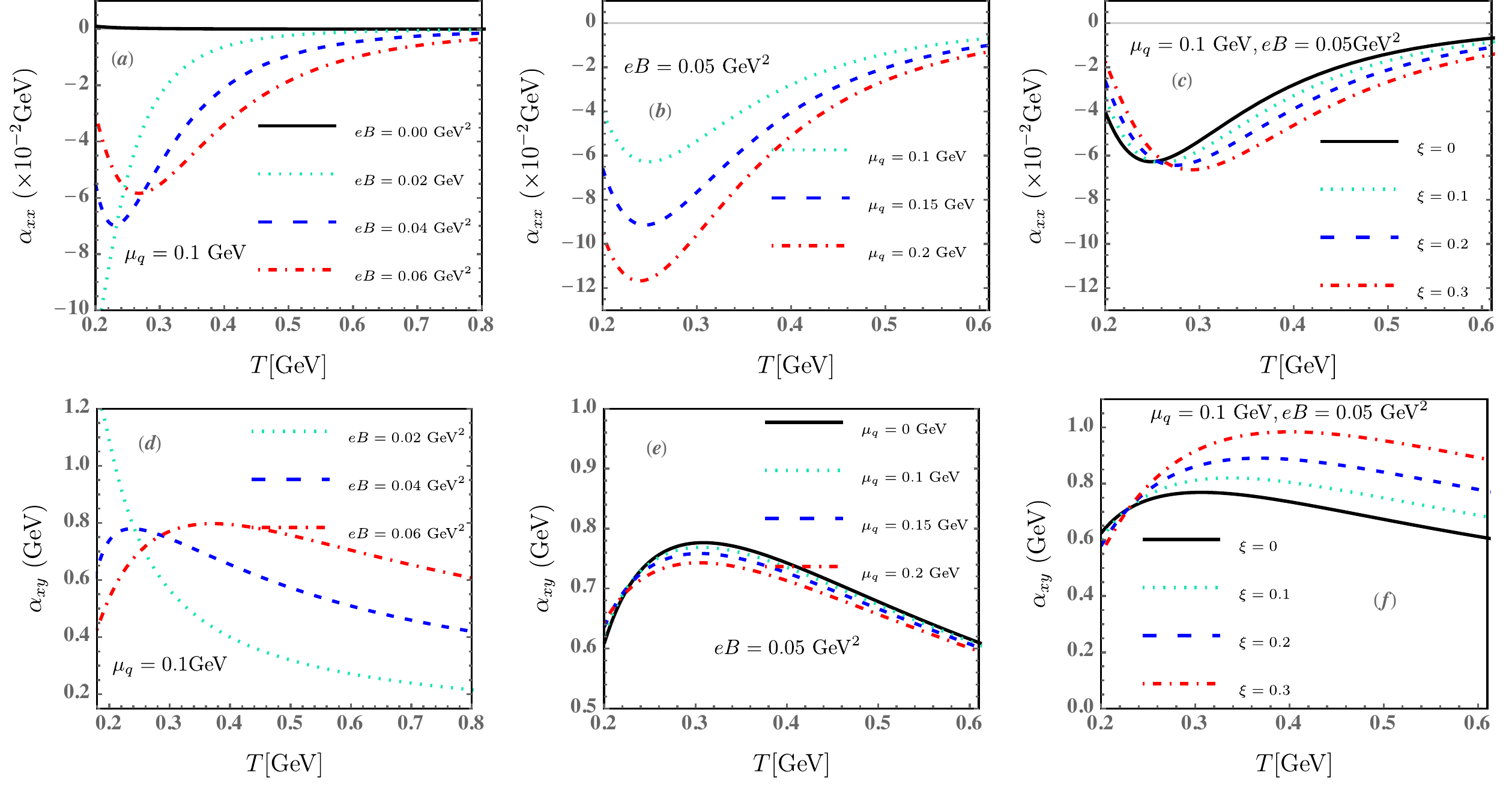}
      	\caption{\label{fig:alpha}   [Diagrams (a) and (d)] The temperature dependences of the thermoelectric conductivity ($\alpha_{xx}$) and the Hall-like thermoelectric conductivity  ($\alpha_{xy}$)  for $\mu_{q}=0.1$~GeV at different magnetic fields, namely, $eB=$ 0~$\mathrm{GeV^{2}}$ (black solid line), 0.02~$\mathrm{GeV^{2}}$ (cyan dotted lines), 0.04~$\mathrm{GeV^{2}}$ (blue dashed lines), 0.06~$\mathrm{GeV^{2}}$ (red dotted-dashed lines). [Diagrams (b) and (e)] The temperature dependences of $\alpha_{xx}$  and $\alpha_{xy}$ for $eB=0.05~\mathrm{GeV^{2}}$ at  $\mu_{q}=$ 0~GeV (black solid line), 0.1~GeV (cyan dotted lines), 0.15~GeV (blue dashed lines) and 0.2~GeV (red dotted-dashed lines).  [Diagrams (c) and (f)] The temperature dependences of $\alpha_{xx}$ and $\alpha_{xy}$ for $\mu_{q}=0.1~$GeV and $eB=0.05~\mathrm{GeV^2}$ in a weakly anisotropic QGP  induced by initial spatial expansion with $\xi=0$ (black solid lines), 0.1 (cyan dotted lines), 0.2 (blue dashed lines), 0.3 (red dotted-dashed lines).}
      \end{figure*}
      For the  total   thermoelectric conductivity ($\alpha_{xx}$), the computation is also  limited to nonzero chemical potential case.  From a quantitative respect,  the  first term  in Eq.~(\ref{eq:alpha-xx-xy-0})  is numerically larger than the second term. This  mathematical difference  arises from the different power of momentum in respective integrands. From a qualitative respect,
        for  the  thermoelectric conductivity  of $f$-th flavor quarks,  $\alpha_{xx,q_f}$, its  thermal behavior under  nonzero $B$ and  nonzero $\mu_{q}$ is determined by $\frac{q_f\tau_{q_f}}{1+(q_{f}\omega_{c,q_f}\tau_{q_f})^2}$ and $\bar{f}^0_{q_f}$ in Eq.~(\ref{eq:alpha-xx-xy-0}),  as well as its sign depends on the quark type.
      Thus the absolute  $\alpha_{xx,q_f}$ for various flavors is similar  to  $\sigma_{xx,q_f}$ in  the qualitative and quantitative.
     However, because of the sensitivity of $\alpha_{xx,q_f}$  in charge characteristic and mass,  the qualitative and quantitative behavior of  the total  thermoelectric conductivity  ($\alpha_{xx}$) is different to that of the total electrical conductivity  ($\sigma_{xx}$). 
    In Fig~\ref{fig:alpha} (a),     $\alpha_{xx}$ at a nonzero $B$  exhibits  negative in sign \footnote { In this paper, the sign of $\alpha_{xx}$ for a fixed $\mu_{q}$ at the  vanishing magnetic field exhibits positive.  However, we do not  discuss much  $\alpha_{xx}$ in the  vanishing magnetic field qualitatively or qu antitatively due to it's marginal results ($10^{-4}\sim 10^{-5}$) at $0.2~\mathrm{GeV}<T<0.4~\mathrm{GeV}$.}. 
    At high enough temperature, $\mu_q/T\sim 0$, the Boltzmann factor $e^{\mu_{q}/T}$ in $\bar{f}^0_{q_f}$ becomes  smaller, the numerical difference between $\alpha_{xx,d}+\alpha_{xx,s}$ and $\alpha_{xx,u}$ reduces gradually.
     As a result,  $\alpha_{xx}$  is closer to zero with increasing $T$, as illustrated in Fig.~\ref{fig:alpha} (a). 
     And with the increase of  $B$, we clearly see that the thermal behavior of $\alpha_{xx,q_f}$ exhibit a valley structure.
    The  $\mu_{q}$  and $\xi$ dependence on the absolute  $\alpha_{xx}$ in  the entire $T$ domain of interest is similar to the counterpart on $\sigma_{xx}$. Furthermore, as $\xi$ grows,  the  minimum of $\alpha_{xx}$ shifts towards higher temperature.
   
      For the total Hall-like thermoelectric conductivity ($\alpha_{xy}$), the estimation is limited to the case of nonzero $B$, which can be easily understood from the corresponding expression in Eq.~(\ref{eq:alpha-xx-xy-0}).
      Similar to $\alpha_{xx}$, the numerical value of $\alpha_{xy}$ for a fixed $\mu_{q}$  is also mainly determined by the first term  in Eq.~(\ref{eq:alpha-xx-xy-0}), the difference is that the sign of  $\alpha_{xy}$ is independent of quark type and always remains positive.
      Since the  thermal behavior of  $\alpha_{xy,q_f}$ 
      mainly depends on $\frac{q_f\omega_{c,q_f}\tau_{q_f}^2}{1+(\omega_{c,q_{f}}\tau_{q_f})^2}$ and $\bar{f}^0_{q_f}$  in Eq.~(\ref{eq:alpha-xx-xy-0}), the  temperature and magnetic field  dependences of $\alpha_{xy}$  are  allied to that of  $\sigma_{xy}$, as shown in Fig.~\ref{fig:alpha}(d).
      In Fig.~\ref{fig:alpha}(e), we clearly  observe that  $\alpha_{xy}$ decreases as $\mu_{q}$  increases at small $T$, which is opposite to  the behavior of  $\sigma_{xy}(\mu_{q})$. 
     This is attributed that at small $T$ though the first term in the expression of $\alpha_{xy}$  is  much larger than the second term in  magnitude, the increment of the  first term with the increase of $\mu_{q}$ is compensated by the more significant reduction of the second term. At high $T$,  $\alpha_{xy}$ remains almost constant with the variation of  $\mu_{q}$   because  $\mu_q/T$ at high $T$ is  small, the  variation of  $e^{\pm\mu_{q}/T}$ in  distribution functions  is negligible compared to the value of $\alpha_{xy}$ itself.
     In the anisotropic medium, $\frac{q_f\omega_{c,q_f}\tau_{q_f}(\xi)^2}{1+\omega_{c,q_f}^2\tau(\xi)^2}$ for $f$-th quarks in Eq.~(\ref{eq:alpha-xx-xy}) is an increasing function of $\xi$  at any given $T$. In addition, the    variation of the  first term for $\alpha_{xy}$ in Eq.~(\ref{eq:alpha-xx-xy}) with $\xi$  is more greater than the counterpart of the  second term.  Therefore,  $\alpha_{xy}$ increases as $\xi$ increases in  the entire $T$ domain of interest, as shown in Fig.~\ref{fig:alpha}(f).
 
  \begin{figure*}
	\includegraphics[width=7.in,height=2.in]{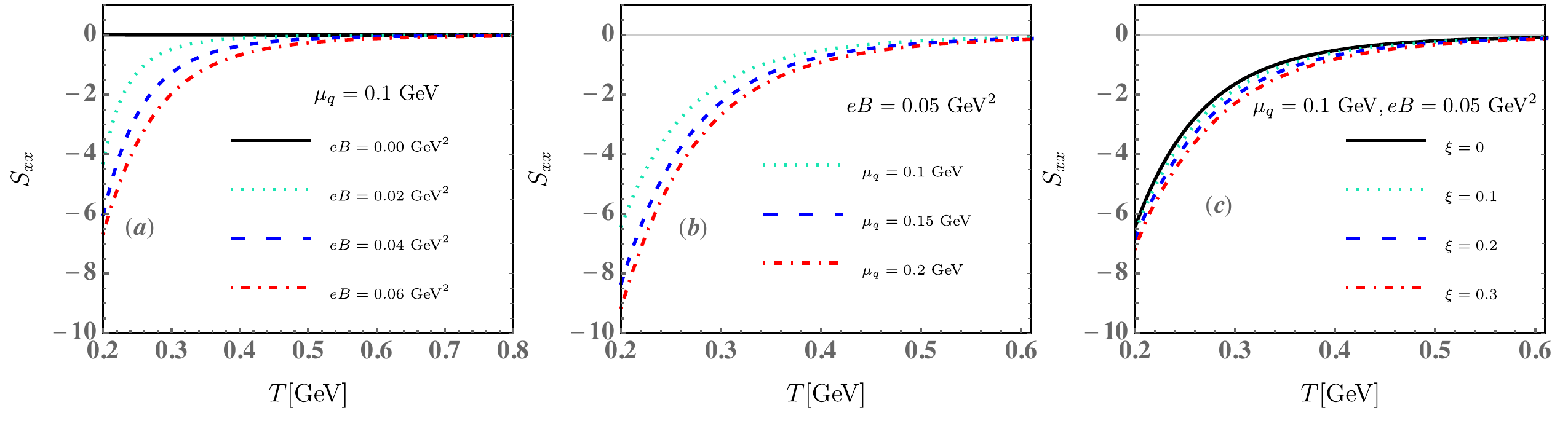}
	\caption{\label{fig:seebeck} [Diagram (a)] The temperature dependence of  the Seebeck coefficient ($S_{xx}$)  for $\mu_{q}=0.1$~GeV at different magnetic fields, namely $eB=$ 0~$\mathrm{GeV^{2}}$ (black solid line), 0.02~$\mathrm{GeV^{2}}$ (cyan dotted line), 0.04~$\mathrm{GeV^{2}}$ (blue dashed line), 0.06~$\mathrm{GeV^{2}}$ (red dotted-dashed line). [Diagram (b)] The temperature dependence of $S_{xx}$ for $eB=0.05~\mathrm{GeV^{2}}$ at  $\mu_{q}=$  0.1~GeV (cyan dotted line), 0.15~GeV (blue dashed line) and 0.2~GeV (red dotted-dashed line).  [Diagram (c)] The temperature dependence of $S_{xx}$ for $\mu_{q}=0.1$~GeV and $eB=0.05~\mathrm{GeV^2}$ in a weakly anisotropic medium with $\xi=0$ (black solid line), 0.1 (cyan dotted line), 0.2 (blue dashed line), 0.3 (red dotted-dashed line). }
\end{figure*}

 \begin{figure*}
	\includegraphics[width=7in,height=2.in]{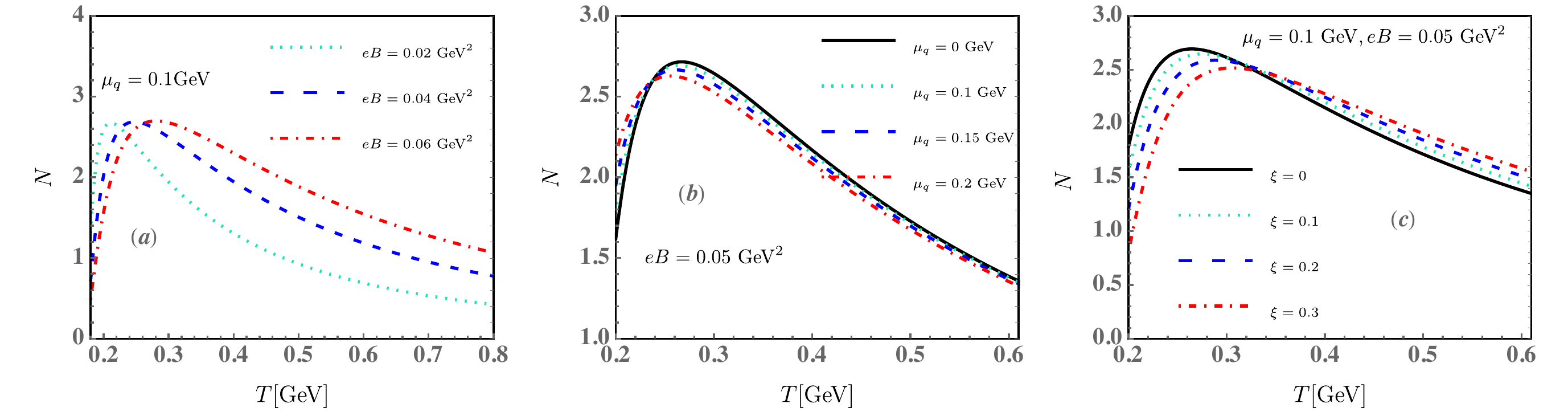}
	\caption{\label{fig:nernst} [Diagram (a)] The temperature dependence of the Nernst signal ($N$)  for $\mu_{q}=0.1$~GeV at different magnetic fields, namely $eB=$ 0.02~$\mathrm{GeV^{2}}$ (cyan dotted line), 0.04~$\mathrm{GeV^{2}}$ (blue dashed line), 0.06~$\mathrm{GeV^{2}}$ (red dotted-dashed line). [Diagram (b)] The temperature dependence of $N$ for $eB=0.05~\mathrm{GeV^{2}}$ at  $\mu_{q}=$ 0~GeV  (black solid line), 0.1~GeV  (cyan dotted line), 0.15~GeV  (blue dashed line) and 0.2~GeV (red dotted-dashed line).  [Diagram (c)] The temperature dependence of $N$ for $\mu_{q}=0.1~$GeV and $eB=0.05~\mathrm{GeV^2}$ in a weakly anisotropic medium with $\xi=0$ (black solid line), 0.1 (cyan dotted line), 0.2 (blue dashed line), 0.3 (red dotted-dashed line).}
\end{figure*}

   Since the quantitative and qualitative behaviors of both the total Seebeck coefficient ($S_{xx}$) and  the total Nernst signal ($N$) are attributed by  the intricate interplay of four conductivity tensors,  we only  phenomenologically discuss the impacts of magnetic field ($B$), quark chemical potential ($\mu_{q}$), and  anisotropic parameter ($\xi$)  on   $S_{xx}$ and  $N$ step by step.   
    In Fig.~\ref{fig:seebeck} (a), we display the temperature dependence of  $S_{xx}$ for finite $B$ at $\mu_{q}=0.1$~GeV. 
    In the  semiconductor,  a positive (negative) $S_{xx}$ implies that the generated electric current runs toward (away from) the direction of high temperature end in an electron (hole) rich side. In other word, $S_{xx}$ is negative for negatively charge carriers and positive for positively charge carriers.
    Similarly, in  the QGP,  the  sign of  $S_{xx}$  is positive (negative), indicating that  the major carriers who dominate the conversion from a temperature gradient to  an electric field  are positively (negatively) charged  quarks.
    In our work, the sign of $S_{xx}$ in the  QGP  for $\mu_{q}=0.1~$ GeV at zero $B$ is positive and  the numerical values of $S_{xx}$ are in    $0.0006<S_{xx}<0.003$ at $0.2~\mathrm{GeV}\leqslant T\leqslant 0.4~\mathrm{GeV}$. It's worth noting that  our results  are close to  the results  in Ref.~\cite{Bhatt:2018ncr}, where the values of  $S_{xx}$ for the QGP at  the vanishing magnetic field  for $\mu_{B}=0.05~\mathrm{GeV}$ lie in the regime of  $0<S_{xx}<0.005$  under the same temperature region.
    At the nonzero (zero) magnetic field,
     the sign of $S_{xx}$  in the QGP  is   negative (positive), which is consistent with the sign of $\alpha_{xx}$. But  the thermal behavior of   $S_{xx}$ is  monotonic instead of nonmonotonic. 
     At high enough $T$,  $S_{xx}$ approaches to zero, indicating the system along $x$-axis is close to an isothermal state.
    And the absolute value of $S_{xx}$ in the  QGP increase as $B$ increases. From Fig.~\ref{fig:seebeck} (b-c), we observe  that   the absolute $S_{xx}$ significantly  increases as  $\mu_{q}$ and  $\xi$ increase at $T<0.4~\mathrm{GeV}$.
    Next, we start to  discuss the Nernst effect.   For vanishing $B$, there is no Lorentz force to bend the trajectories of the thermally diffusing charge carriers, so the Nernst effect is absent.
    Fig.~\ref{fig:nernst} (a) shows the evolution of  $N$ as a function of $T$ at different $B$ for $\mu_{q}=0.1$ GeV.
    Unlike  $S_{xx}$,   $N$ in sign is independent of the charge carrier type, which can well understand from the associated expression.
   We can clearly see  the value of $N$  always remains positive and  the thermal evolution of $N$ at the magnetic field has  a peak structure. We also observe as $B$  grows,  $N$  decreases at small $T$   whereas increases at high $T$, which is consistent  with  $\sigma_{xy}(B)$ and $\alpha_{xy}(B)$ in the qualitative.
    In this work, the maximum of $N$ for  $\mu_{q}=0.1~\mathrm{GeV}$ approximately is  2.8.   We  can zoom in Fig.~\ref{fig:nernst} (b)  and observe  that the  $\mu_{q}$  dependence of  $N$  is very similar to  $\alpha_{xy}(\mu_q)$, except that $N$ weakly decreases as $\mu_{q}$ increases at $T<0.26~\mathrm{GeV}$. Anyway, the effect of $\mu_{q}$ on $N$ is not obvious.
   The effect of momentum anisotropy induced by initial spatial expansion on $N$ at $\mu_{q}=0.1$~GeV for $eB=0.05~\mathrm{GeV^2}$ is plotted in Fig.~\ref{fig:nernst} (c). We see  that as $\xi$ increases, $N$  decreases at relatively low temperature  then increases at high temperature, which is  qualitatively  akin   to $\sigma_{xy}(\xi)$. Furthermore, with the rise in  $\xi$, the maximum of  $N$ falls and slightly shifts to higher temperature.

      \begin{figure*}
     	\includegraphics[width=3in,height=2.5in]{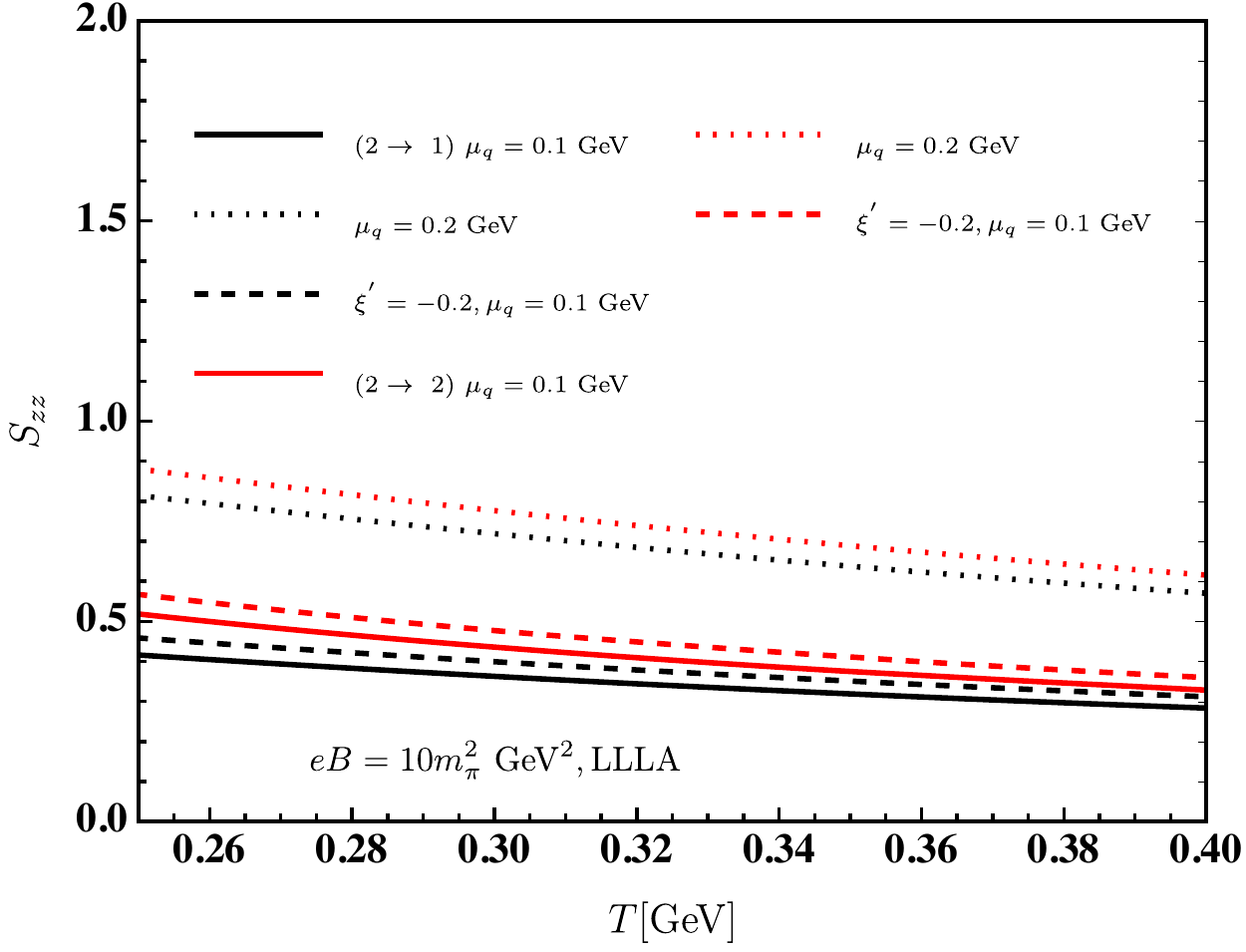}
     	\caption{\label{fig:LLL} The temperature  dependence of  Seebeck coefficient along  the direction of magnetic field  ($S_{zz}$) for 2$\rightarrow$1 process (black lines) and 2$\rightarrow2$ process (red lines) in the  LLL approximation. Solid lines and dotted lines represent the calculations of $S_{xx}$ for $eB=10m_{\pi}^2~\mathrm{GeV^{2}}$ are performed at $\mu_{q}=0.1$~GeV  and at 0.2 ~GeV, respectively. Dashed lines denote the temperature dependence of $S_{xx}$ at $\mu_{q}=0.2~$GeV  for $eB=10m_{\pi}^2~\mathrm{GeV^{2}}$ in an  anisotropic medium induced by strong magnetic field (we take $\xi'=-0.2$).
     		Within the assumed regime  $m_q^2\ll\alpha_{s,B}eB\lesssim T^2\lesssim eB$, we  artifically take $eB=10m_{\pi}^2~\mathrm{GeV^{2}}$ and the appropriate temperature  can be concentrated within the region   $0.25~\mathrm{GeV}\leqslant T\leqslant 0.4~\mathrm{GeV}$.  }.  
     \end{figure*}
     
\begin{figure*}
	\includegraphics[width=6.8in,height=2.in]{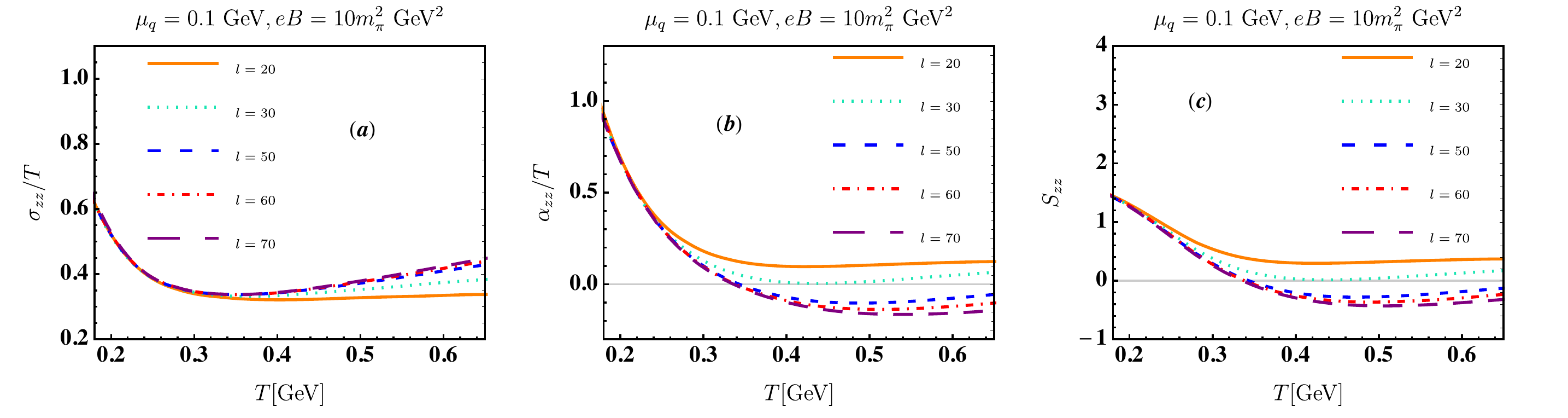}
	\caption{\label{fig:seezz} The temperature dependences of the  scaled  longitudinal electrical  conductivity  $(\sigma_{zz}/T)$ [in diagram (a)], the scaled longitudinal thermoelectric conductivity $(\alpha_{zz}/T)$ [in diagram (b)] and  the longitudinal Seebeck coefficient  ($S_{zz}$) [in diagram (c)] within different Landau levels ($l=$20 (orange solid lines),~30 (cyan dotted lines),~50 (blue dashed lines),~60 (red dotted-dashed lines) and~70 (purple wide-dashed lines)) for $eB=10m_{\pi}^2~\mathrm{GeV^{2}}$  at $\mu_{q}=0.1$~GeV. }
\end{figure*}

    The  investigation of  the thermoelectric coefficient   is also  converted  to  the  strong  magnetic background   field. 
     The  calculation of  the longitudinal conductivity tensors   is  first performed under the   LLL approximation.
     In the  LLL approximation, $2\rightarrow1$ scattering process and $2\rightarrow2$ scattering  process  are  taken into account. Although the inverse  relaxation times in Eq.~(\ref{eq:tau-21-LLL}) and Eq.~(\ref{eq:tau-22-LLL})  for two kind of processes are  of the same order of $\alpha_{s,B}$, 
      the numerical value of inverse relaxation time for $2\rightarrow2$ process is marginal  compared with that for $2\rightarrow1$ process.
     Since the largest inverse relaxation time determines the final inverse relaxation time,  the $2\rightarrow1$ process is significantly dominated  over  $2\rightarrow2$  process in the  strong magnetized QGP. Nevertheless, we still can compute the respective contribution to the component of the longitudinal Seebeck coefficient, namely, $S_{zz}$, in the  LLL approximation.  
     Fig.~\ref{fig:LLL}  demonstrates  that $S_{zz}$ of isotropic QGP in the  LLL approximation for $\mu_{q}=0.1~\mathrm{GeV}$  decreases with increasing temperature. Unlike $S_{xx}$ at weak magnetic field, the sign of $S_{zz}$ at strong magnetic field within the  LLL approximation is positive, indicating  that  the dominant charge carriers  for converting the thermal gradient along the direction of magnetic field to electric field are positively charged quarks. Yet, in Ref.~\cite{Dey:2020sbm} the result  of $S_{zz}$ is negative because the sensitivity  of the  relaxation time in the quark chemical potential is not taken into account.
     We also observe the value of  $S_{zz}$ for 2~$\rightarrow$~1 process is comparable with that for 2~$\rightarrow$~2 process. Actually,  $S_{zz}$  in  the  LLL approximation is  independent of $B$  due to the fact that  the $B$-dependent factors   in  the numerator of $S_{zz}$  and the counterparts in the denominator  cancel out. Similar to the previous  calculation in weak magnetic field, we also consider the  effect of quark chemical potential ($\mu_{q}$)  on $S_{zz}$ in the  LLL approximation. As illustrated in Fig.~\ref{fig:LLL},  $S_{zz}$ for both  2~$\rightarrow$~1 process and 2~$\rightarrow$~2 process  numerically  increases as $\mu_{q}$ increases in entire $T$ domain of interest.
     In comparison to the isotropic medium,   $S_{zz}$ in a strong magnetic field-driven anisotropic medium (we take~$\xi'=-0.2$) remains unchanged in the qualitative but has a quantitative  enhancement.
     Finally, the calculation of $S_{zz}$ is also extended to  a more realistic regime  $g^2T^2\ll eB$ in which the contribution from higher Landau levels (hLLs)  are considered.
    In Fig.~\ref{fig:seezz} (a) and (b),  as  Landau level ($l$) increases, we note that   the scaled longitudinal electrical conductivity   ($\sigma_{zz}/T$) increases, whereas  the scaled longitudinal  thermoelectric conductivity ($\alpha_{zz}/T$)  decreases.  In our work, the numerical values  of $\sigma_{zz}/T$  at $eB=10m_{\pi}^2~\mathrm{GeV^{2}}$ and $\mu_{q}=0~\mathrm{GeV}$ within the consideration of hLLs  contribution  are in $0.3<\sigma_{zz}/T<0.7$ for $0.2~\mathrm{GeV}\leqslant T\leqslant 0.5~\mathrm{GeV}$, which  is  consistent with the result ($0.2<\sigma_{zz}/T<0.7$) of existing report~\cite{Kurian:2017yxj}  in the same configuration. And in present work, $\sigma_{zz}/T$ within the effect of hLLs   also quantitatively lies in the range of  Lattice QCD results ($0.1\leqslant\sigma_{el}/T\leqslant1.0$) from Ref.~\cite{lattice}. 
      In Fig.~\ref{fig:seezz} (c),  we see that the  thermal behavior of  $S_{zz}$ beyond the LLL approximation with $l=20$ is similar to the counterpart of $S_{zz}$ in the  LLL approximation.
         Due to  the decreasing features of  both $\alpha_{zz}$ and $1/\sigma_{zz}$ with  increasing  Landau level  as shown in Fig.\ref{fig:seezz}(a-b), the value of  $S_{zz}$ can be  suppressed as  the Landau level rises. And  $S_{zz}$  even changes  sign  from positive to negative as the Landau level grows, which mimics  the dependence of $\alpha_{zz}$ on the Landau level. The change of $S_{zz}$ in sign  reflects the dominant charge carriers for  converting the  heat  gradient along $z$-axis to electric field become negatively charged quarks rather than positively  charged quarks as the increase of Landau level. Furthermore, at higher temperature,   $S_{zz}$ for various Landau levels converges to zero, indicating that the system along $z$-axis is in a nearly  isothermal state. 

\section{conclusion}\label{sec:summary}
    A theoretical investigation  on  the  Seebeck effect  and the Nernst effect of  QGP  in the magnetic fields has been presented. The associated  Seebeck  and Nernst signal are the functions regarding the  electrical conductivity tensors and the thermoelectric conductivity tensors, which can be obtained by  solving the relativistic Boltzmann  equation  under  the relaxation time approximation. 
     We found  in the presence (absence) of weak magnetic field  along $z$-axis,  $S_{xx}$ for a nonzero  quark chemical potential  is negative (positive) in sign, indicating that the dominant charge carriers for converting  heat gradient into electric field are negatively (positively) charged  quarks. We  found as temperature increases,  $S_{xx}$   first decreases (absolute value increases)  then gradually tends to zero, which implies that at high enough temperature  the system  reaches an isothermal state. And the absolute value of  $S_{xx}$ has a further enhancement with the increase in magnetic field and quark chemical potential. 
     We also extended the exploration to an anisotropic QGP, where  the partons exhibit a local anisotropy ($\xi\neq0$) in the momentum space due to the  rapid expansion of initial fireball along the beam direction.
   	 The results showed that   the absolute value of  $S_{xx}$  in a weakly anisotropic medium   has an obvious enhancement  compared to that in  an isotropic medium  ($\xi=0$), and with  the increase of $\xi$ this increment can be strengthened.
    	Different from monotonous thermal behavior of $S_{xx}$, the temperature dependence  of  the Nernst signal ($N$)  for  weak magnetic field exhibits a peak structure, and $N$ in sign  is independent of the type of  charge carriers.   As magnetic field as well as anisotropic parameter ($\xi$) increase, $N$ decreases at relatively small temperature  whereas  increases at high temperature.
      In contrast to the effects of magnetic field and momentum anisotropy, with the increase in quark chemical potential,  $N$ increases at small temperature  whereas  decreases  at high temperature.
    
	 In  the  strong magnetic field, the  Seebeck coefficient along the direction of magnetic field, $S_{zz}$,  has been calculated under the  LLL approximation and beyond  the LLL approximation. The value of $S_{zz}$ in the  LLL approximation  always remains positive and increases as quark chemical potential rises. And we  found $S_{zz}$ in the LLL approximation  is  independent of the magnetic field strength. Under the  same condition, the value of $S_{zz}$ for $2\rightarrow2$ process is comparable with that for $2\rightarrow1$ process, even though the former process  is far less important than the latter process in the strongly magnetized QGP.  In addition, $S_{zz}$ in the  anisotropic QGP  induced by strong magnetic field with $\xi'=-0.2$ has an overall enhancement compared to that   in the isotropic medium.   With the increase of  Landau level, $S_{zz}$ decreases  and even  changes the sign   from positive to negative. 
      For the  future investigation,  we  may  study the Seebeck coefficient  and  the Nernst signal at the  magnetic field in the hadronic phase and near the transition phase region  based on van der Waals hadron resonance gas  (VDWHRG) model and QCD effective models (e.g., Polyakov Nambu-Jona-Lasinio model and Polyakov Quark Meson model), respectively. Especially, a direct comparison of the results in  the hadronic and partonic phases would be instructive.
      
	\section*{ACKNOWLEDGMENTS}
    We thank  Arpan Das for bringing some recent literature to our attention.
    This research is supported   by the National Natural Science Foundation of China under Grant No.11935007,
    Guangdong Major Project of Basic and Applied Basic Research No. 2020B0301030008,  and   the  Fundamental Research Funds for the Central Universities under Grant No.2020CXZZ107.
   \appendix
   \section{}
   The collision term of species $a_1$  for the  binary process $a_1(P_1)+a_2(P_2)\rightarrow a_3(P_3)+a_4(P_4)$ is given by~\cite{RKT,Weldon:1982aq}
   \begin{eqnarray}
   C[f_1]=&&\sum_{pro}\frac{d_2}{1+\delta_{34}}\frac{1}{2\epsilon_{1}}\prod_{i=2}^{4}\int d\Gamma_i(2\pi)^4
   \delta^4(P_{tot})
   |\mathcal{M}^{pro}_{}|^2\nonumber\\
   &&\times\left[f_1f_2(1\pm f_3)(1\pm f_4)-f_3f_4(1\pm f_1)(1\pm f_2)\right]. \nonumber\\
   \end{eqnarray}
    In the above, $P_{tot}=P_1+P_2-P_3-P_4$. $P_{i=1,2,3,4}=(\epsilon_{i},\mathbf{p}_i)$ denotes the four-momentum of particle, where $\epsilon_{i}=\sqrt{\mathbf{p}^2_i+m_i^2}$. 
      We  use a notation $d\Gamma_i=\frac{d^3\mathbf{p}_{i}}{(2\pi)^32\epsilon_{i}}$ for convenience. 
      The factor $1/(1+\delta_{34})$ is introduced to avoid  double counting when particle $a_1$ and $a_2$ are identical. 
      $\mathcal{M}^{pro}$ is the scattering amplitude for a specific binary process.
   Considering  the distribution  slightly derivate the equilibrium, hence  the local momentum distribution  of $i$-th species is given by  $f_i=\bar{f}_{i}^0+\delta f_{i}=\bar{f}_{i}^0+\beta\bar{f}_{i}^0(1\pm \bar{f}^0_{i})\chi_{i}(\mathbf{p}_i)$, 
 where $\chi_i$ is the response function in the effect of electric field. For (anti-)quarks, the associated response functions hold the relation of  $\chi_{\bar{q}}(\mathbf{p})=\chi_{q}(-\mathbf{p})$ due to the charge conjugation symmetry. However, for gluons $\chi_{g}$  is zero.  Using the  detailed balance condition $\bar{f}_1^0\bar{f}^0_2(1\pm \bar{f}^0_3)(1\pm \bar{f}^0_4)=\bar{f}_3^0\bar{f}_4^0(1\pm \bar{f}^0_1)(1\pm \bar{f}^0_2)$, 
    the collision term can be rewritten as
   \begin{eqnarray}
    C[f_1]=&&\sum_{pro}\frac{d_2}{1+\delta_{34}}\frac{1}{2\epsilon_{\mathbf{p}_1}}\prod_{i=2}^{4}\int d\Gamma_i(2\pi)^4
    \delta^4(P_{tot}4)
    |\mathcal{M}^{pro}_{}|^2\nonumber\\
    &&\times[\bar{f}^0_1\bar{f}^0_2(1\pm \bar{f}^0_3)(1\pm \bar{f}^0_4)]\beta\nonumber\\
   &&\times(\chi_3(\mathbf{p}_3)+\chi_4(\mathbf{p}_4)-\chi_2(\mathbf{p}_2)-\chi_1(\mathbf{p}_1)).
   \end{eqnarray}
   Since the response function $\chi_i(\mathbf{p}_i)$ is an odd function of $\mathbf{p}_i$ for (anti-)quarks or is zero for gluons, whereas other integrand is an even function of $\mathbf{p}_i$, the result of  the integral related to $\chi_j(\mathbf{p}_i)$ ($i=2,3,4$)  is  zero.  Using  relaxation time approximation  $C[f_1]=-\delta f_1/\tau_1=-\beta\bar{f}_1^0(1\pm \bar{f}_1^0)\chi_1(\mathbf{p}_1)/\tau_1$,  we can obtain momentum dependent  thermal relaxation time of species $a_1$, namely, 
   \begin{equation}\label{eq:tau1}
   \tau_{1}^{-1}=\sum_{pro}\frac{d_2}{\delta_{34}+1}\int\frac{d^3\mathbf{p}_2}{(2\pi)^3}\frac{\bar{f}_{2}^0(1\pm\bar{f_3^0})(1\pm\bar{f}_4^0)}{(1\pm \bar{f}_1^0)}\int dt \frac{d\sigma^{pro}}{dt},
   \end{equation}
   where  $\frac{d\sigma^{pro}}{dt}=\frac{\langle|M^{pro}|^2\rangle}{16\pi s^2}$ denotes the differential scattering  cross section with respect to  the Mandelstam variables $s$, $t$, $u$.

  Due to the large angle scattering is the most efficient mechanism for the transport process in  a plasma with long-range interaction~\cite{transport-rate,zhuang:1995,Thoma:1993vs},  a phenomenological weight factor  $\sin \theta^2/2=2tu/s^2$ ($\theta$ is the scattering angle in the center of mass system) is introduced in Eq.~(\ref{eq:tau1}).
    Alternatively, when the momentum transfer  $|\mathbf{p}_1-\mathbf{p}_3|=|\mathbf{p}_2-\mathbf{p}_4|$ is small or $t=|P_1-P_3|=|P_2-P_4|\ll \sqrt{s}$,  we assume that $\bar{f}_1=\bar{f}_3$ and  $\bar{f}_2=\bar{f}_4$ for the elastic scatterings,   the thermal relaxation time can finally be rewritten as 
   \begin{equation}\label{eq:tau}
   \tau_{1}^{-1}=\sum_{pro}\frac{d_2}{\delta_{34}+1}\int\frac{d^3\mathbf{p}_2}{(2\pi)^3}\bar{f}^0_{2}(1\pm\bar{f}^0_4)\int dt \frac{d\sigma^{pro}}{dt}\frac{2tu}{s^2}.
   \end{equation}
    The integration  in terms of  $t$-channel  only has the  logarithmic infrared divergence~\cite{Thoma:1993vs}.  And this divergence  can be regulated by  restricting the $t$-channel integration from $-s$ to $-\mu_{D}^2$, 
   where  $\mu_{D}^2=g^2T^2$  is the infrared regulator in the upper bound of the $t$ integration~\cite{Thoma:1993vs}.
 \section{}
  Apart from  $q+\bar{q}\rightarrow g$ process,  another dominant process in the  LLL approximation with the specfic regime   $m_q^2\ll \alpha_{s,B}eB\ll T^2\ll eB$ is $t$ channel  $q~(P)+\bar{q}~(P'')\rightarrow q~ (P')+\bar{q}~(P''')$ scattering. The associated collision  term   has been  presented by K.Hattori $et$ $al$ in Ref.~\cite{Hattori:2016lqx} using the leading order perturbative QCD approach. 
   	We extend their result to
   	  nonzero quark chemical potential case, which 
   	  has the following form
   	  \begin{widetext}
   \begin{eqnarray}\label{eq:2-2 process}
   C[f_{B,q}(p_z,\mu_{q})]_{2\rightarrow2}=&&2g^4T_{R}C_{R}(\frac{|q_{f}eB|}{2\pi})(16m_f^4)\frac{1}{(2\epsilon_{f,0})^2}\beta\bar{f}^0_{B,q}(p_z,\mu_{q})(1-\bar{f}^0_{B,\bar{q}}(p_z,\mu_{\bar{q}}))\nonumber\\
   &&\times\int\frac{dp_{z}'}{2\pi}\frac{1}{(\epsilon'_{f,0})^2}\frac{\epsilon_{f,0}\epsilon_{f,0}^{'}}{|\epsilon_{f,0}p_z'-\epsilon'_{f,0}p_z|}\int\frac{d^2\mathbf{q}_{\perp}}{(2\pi)^2}e^{-\frac{\mathbf{q}_{\perp}^2}{eB}}\frac{1}{\left(q^2+\mathrm{Re}\Pi_z(q)+i\mathrm{Im}\Pi_z(q)\right )^2}\nonumber\\
   &&\times\bar{f}^0_{B,\bar{q}}(p'_z,\mu_{\bar{q}})(1-\bar{f}^0_{B,q}(p'_z,\mu_{q}))(\chi_q(p_z')-\chi_q(p_z)),
   \end{eqnarray}
    \end{widetext}
   where $\mathbf{q}=\mathbf{p'}-\mathbf{p}=\mathbf{p''}-\mathbf{p'''}$ is the  momentum transfer, $q^2=\mathbf{q}_{\perp}^2-q_{||}^2$ and  $-q_{||}^2=-(P'-P)_{||}^2=2(\epsilon_{f,0}\epsilon'_{f,0}-p_{z}p_{z}'-m_{f}^2)$. $\epsilon_{f,0}=\sqrt{p_z^2+m_f^2}$ and $\epsilon'_{f,0}=\sqrt{p_z'^2+m_f^2}$.  $\mathrm{Re}\Pi_z(q)$ and $\mathrm{Im}\Pi_z(q))$ in Eq.~(\ref{eq:2-2 process}) are the real and imaginary parts of gluon self-energy along the direction of magnetic field, respectively. In the static limit ($q_0\rightarrow0$), $\mathrm{Re\Pi_z }$ is the Debye mass $m_{D,B}$.  In the  work of  M. Hasan $et$ $al$ ~\cite{Hasan:2018kvx}, the  imaginary part of gluon self-energy in  the strong magnetic field is given as
   \begin{eqnarray}
   \frac{\mathrm{Im}\Pi_z(\mathbf{q})}{q_0}\bigg|_{q_0\rightarrow0}=-\frac{g^2}{\mathbf{q}}\frac{\pi T^2}{2}-\frac{g^2}{q_z^2}\frac{\sum_{f}m_f^2|q_feB|}{8\pi T}.
   \end{eqnarray}
   Thus  we can note that $\mathrm{Im}\Pi_z(\mathbf{q})$ vanishes in the static limit. Furthermore,  in chirality nonflip case $p_z\cdot p_z'>0$, $|\epsilon_{f,0}p_z'-\epsilon'_{f,0}p_z|$ and $-q_{||}^2$ can rewritten as~\cite{Hattori:2016lqx},
   \begin{eqnarray}
   |\epsilon_{f,0}p_z'-\epsilon'_{f,0}p_z|=\frac{m_f^2|p_z'^2-p_z^2|}{|\epsilon_{f,0}p_z'+\epsilon'_{f,0}p_z|}\approx\frac{m_f^2}{\epsilon_{f,0}}|p_z'-p_z|
    \end{eqnarray}
    and
     \begin{eqnarray}
   -q_{||}^2&=&2(\epsilon_{f,0}\epsilon'_{f,0}-p_{z}p_{z}'-m_{f}^2)=\frac{2m_f^2(p_z'-p_z)^2}{\epsilon_{f,0}\epsilon'_{f,0}+p_{z}p_{z}'+m_{f}^2}\nonumber\\
  && \approx\frac{m_f^2}{\epsilon_{f,0}^2}(p_{z}'-p_z)^2,
   \end{eqnarray}
   respectively.
   In the hierarchy of scale $eB\gg T^2$, the form factor $e^{-\frac{\mathbf{q}^2_\perp}{eB}}$ can reasonably be  neglected  due to   $\mathbf{q}^2_\perp\sim -q_{||}^2\lesssim T^2\ll eB$, and  $m_{D,B}^2$ always dominates over $q_{||}^2$ in the regime  $\alpha_{s,B}eB\gg m_q^2$. Therefore,  Eq.~(\ref{eq:2-2 process}) can further reduce to 
\begin{widetext}
   \begin{eqnarray}
   C[f_{B,q}(p_z,\mu_{q})]_{2\rightarrow2}&=&8\pi\alpha_{s}^2T_RC_R(\frac{eB}{2\pi})\frac{m_{f}^2\beta}{\epsilon_{f,0}}\bar{f}^{0}_{B,q}(p_z,\mu_{q})\left(1-\bar{f}^{0}_{B,\bar{q}}(p_z,\mu_{\bar{q}})\right)\int\frac{dp_{z}'}{2\pi}\frac{1}{|p_{z}'-p_{z}|\epsilon'_{f,0}/\epsilon_{f,0}}\\\nonumber
   &&\times\frac{1}{m_{D,B}^2}\bar{f}^{0}_{B,\bar{q}}(p'_z,\mu_{\bar{q}})\left(1-\bar{f}^{0}_{B,q}(p'_z,\mu_{q})\right)
   \left(\chi_q(p_z')-\chi_q(p_z)\right ).
   \end{eqnarray}
\end{widetext}
   For the small $p_z-p_z'$, $\chi_q(p_z')-\chi_q(p_z)$ can be approximated as 
   \begin{eqnarray}
   (\chi_q(p_z')-\chi_q(p_z))&\approx&(p_z'-p_z)\partial_{p_z}\chi_q(p_z)\nonumber\\
 & =&-(p_z'-p_z)\partial_{p_z'}\chi_q(p_z').
   \end{eqnarray}
 Therefore, the collision term for $2\rightarrow2$ process when $m_q^2\ll\alpha_seB$ is given by 
   \begin{eqnarray}
   C[f_{B,q}(p_z,\mu_{q})]_{2\rightarrow2}&=&-2\alpha_{s,B}^2T_RC_R(\frac{eB}{\pi})\frac{m_{f}^2\beta}{E_{f,0
   	}m_{D,B}^2}\nonumber\\
   &&\times\bar{f}^{0}_{B,q}(p_z,\mu_{q})\left(1-\bar{f}^{0}_{B,q}(p_z,\mu_{q})\right)\nonumber\\
 &&\times\bar{f}_{B,\bar{q}}^{0}(p_z,\mu_{\bar{q}})\left(1-\bar{f}_{B,\bar{q}}^{0}(p_z,\mu_{\bar{q}})\right)\nonumber\\
 &&\times\chi_q(p_z).
   \end{eqnarray}
    In the LLL approximation with the hierarchy  of scale $eB\gg T^2$, $m_{D,B}^2\approx \sum_{f}\frac{\alpha_{s,B}|q_{f}eB|}{\pi}=2\alpha_{s,B}T_RC_{R}\left(\frac{eB}{\pi}\right)$. 
  Using $C[f_{B,q}]=-\beta\bar{f}_{
  	B,q}^0(1+\bar{f}^0_{B,q})\chi_q/\tau_{B,q}$,
    we finally get thermal  relaxation time of (anti-)quarks  for $f$-th flavor for 2$\rightarrow$2 process,
   \begin{equation}\label{eq:B8}
   \frac{1}{\tau_{B,q(\bar{q})}}\bigg|_{2\rightarrow2}
   =\alpha_{s,B}\frac{m_{q}^2}{\epsilon_{f,0
   	}}\bar{f}^{0}_{\bar{q}(q)}(p_z,\mu_{\bar{q}(q)})\left(1-\bar{f}^{0}_{\bar{q}(q)}(p_z,\mu_{\bar{q}(q)})\right).
   \end{equation}

  In the  anisotropic medium induced by strong magnetic field,  the thermal relaxation time associated with the anisotropic parameter ($\xi'$) can be obtained by straightforwardly  substituting $\bar{f}_{B,q}^0$ and $\alpha_{s,B}$ in Eq.~(\ref{eq:B8}) with $\bar{f}_{B,q}^{\xi'}$ and   $\alpha_{eff,B}$, respectively.

\end{document}